\documentclass[aps,prd,preprint,superscriptaddress,nofootinbib,longbibliography]{revtex4-1}
\pdfoutput=1

\usepackage[T1]{fontenc}

\usepackage{amsmath,amssymb,mathtools}
\usepackage{bbm}                
\usepackage{bm}                 
\usepackage{slashed}            

\usepackage{graphicx}
\usepackage{booktabs}
\usepackage{array}
\usepackage{multirow}
\usepackage{enumitem}           
\usepackage{siunitx}

\usepackage[hidelinks]{hyperref}
\usepackage[capitalise]{cleveref}

\usepackage{xcolor}
\usepackage[normalem]{ulem}     

\newcommand{\mzero}{m_0}
\newcommand{\mhalf}{M_{1/2}}
\newcommand{\MHC}{M_{H_C}}
\newcommand{\MGUT}{M_{\rm GUT}}
\newcommand{\tb}{\tan\beta}
\newcommand{\SO}[1]{SO(#1)}
\newcommand{\SU}[1]{SU(#1)}
\newcommand{\rep}[1]{\mathbf{#1}}

\newcommand{\Xt}{X_t}
\newcommand{\sigSI}{\sigma_{\rm SI}}
\newcommand{\Ohsq}{\Omega h^2}
\newcommand{\mchi}{m_{\tilde\chi}}

\begin{document}

\title{Updated analysis of minimal supersymmetric $\SO{10}$ with a universal soft spectrum}

\author{Cheng-Wei Chiang}
\email{chengwei@phys.ntu.edu.tw}
\affiliation{Department of Physics and Center for Theoretical Physics, National Taiwan University, 
Taipei, Taiwan 10617, R.O.C.}
\affiliation{Physics Division, National Center for Theoretical Sciences,
Taipei, Taiwan 10617, R.O.C.}

\author{Takeshi Fukuyama}
\email{fukuyama@rcnp.osaka-u.ac.jp}
\affiliation{Research Center for Nuclear Physics (RCNP), Osaka University, Ibaraki, Osaka 567-0047, Japan}

\begin{abstract}

We update the analysis of minimal supersymmetric $\SO{10}$ grand unification with a universal, constrained-MSSM-like soft-breaking spectrum. A pair of Yukawa matrices in the $\rep{10}\oplus\overline{\rep{126}}$ Higgs sector fixes the charged fermion masses and quark mixing, the baryon number-violating dimension-five operators that mediate proton decay, and the right-handed neutrino Majorana masses of the type-I seesaw. The simultaneously imposed constraints include: (i) gauge coupling unification and vacuum stability; (ii) dimension-five proton decay (the dimension-six gauge mode also computed and negligible), the $125$-GeV Higgs mass, and the Large Electron-Positron Collider (LEP) chargino limit; and (iii) $\mu\to e\gamma$, the muon $g-2$, and electric dipole moments. We perform a global scan over the soft-breaking parameters, complemented by a constrained-$\mu$ cross-check that fixes the electroweakino sector from the GUT-scale boundary conditions through radiative electroweak symmetry breaking. Subject to a single effective colored Higgs scale $\MHC$ for the proton decay normalization (common to unification and decay, though in a general color triplet mass matrix the two can differ) and to a constrained-$\mu$ treatment, what survives is a narrow mini-split region: heavy, multi-tens-of-TeV scalars with $\mzero \gtrsim 11.0$~TeV for a dressing electroweakino mass $m_\chi=1$~TeV and $\tb \lesssim 8.0^{+5.5}_{-2.0}$, the uncertainty being the Higgs mass theory error. We also examine the fate of the lightest supersymmetric particle as a dark matter candidate, in both the universal model and its free-$\mu$ non-universal-Higgs-mass extension. The surviving region is sharply bounded and testable in the coming decade, most directly by Hyper-Kamiokande proton decay and electroweakino searches, with dark matter direct detection and electric dipole moment experiments.

\end{abstract}


\maketitle

\section{Introduction}\label{sec:intro}

Grand unified theories (GUTs) embed the Standard Model (SM) gauge group into a single
simple group, so that the three gauge couplings meet at a common high energy
scale~\cite{Georgi:1974yf}. Of the candidate groups, $\SO{10}$ is the most economical for
matter: a single irreducible $\rep{16}$ accommodates all fifteen SM
Weyl fermions of one generation together with a right-handed neutrino, which
could give rise to Majorana masses for the light neutrinos through the type-I seesaw mechanism. When the model is
made supersymmetric, the superpartners stabilize the electroweak scale against the
GUT scale and, equally important, sharpen the unification: the
Minimal Supersymmetric Standard Model (MSSM) particle content drives the three
running couplings to a strikingly accurate crossing near $2\times10^{16}$~GeV, a
result that historically turned supersymmetric grand unification from an aesthetic
preference into a quantitative success~\cite{Dimopoulos:1981zb,Sakai:1981gr,Dimopoulos:1981yj}.
The most predictive realization of the matter sector couples each $\rep{16}$ to just
two Higgs representations, the $\rep{10}$ and the $\overline{\rep{126}}$, through a
single pair of symmetric Yukawa matrices~\cite{Babu:1992ia,Matsuda:2000zp,Fukuyama:2002ch}.

These Yukawa matrices $(Y_{\rep{10}},Y_{\overline{\rep{126}}})$ control
several observables simultaneously. Once they are fitted to the
measured charged-fermion masses and the Cabibbo--Kobayashi--Maskawa (CKM) 
matrix, the baryon number-violating dimension-five
operators that mediate proton decay in any supersymmetric
GUT~\cite{Weinberg:1981wj,Sakai:1981pk,Dimopoulos:1981dw} are no longer adjustable: the same couplings that reproduce the fermion spectrum also fix the flavor structure of these operators. Through the
$(B\!-\!L)$-breaking vacuum expectation value (VEV) of the $\overline{\rep{126}}$ Higgs, the same pair also fixes the
right-handed-neutrino Majorana masses and hence the type-I seesaw. The proton decay rate
is thus a prediction of the fitted model, up to the overall factor $1/\MHC^2$ set by the effective colored Higgs scale, rather than an adjustable parameter. The dimension-five
operator generated by the exchange of the heavy color triplet Higgs is dressed to a
physical $\Delta B=1$ amplitude by a loop of electroweak gauginos and higgsinos,
whose strength depends on the superpartner spectrum~\cite{Ellis:1981tv,Hisano:1992jj}, so the
nucleon decay rate ties the GUT-scale Yukawa couplings directly to the soft
supersymmetry-breaking masses to be constrained.

Many independent measurements bear on this tightly constrained
model.  They fall into four groups by physical origin: theoretical consistency, including gauge coupling unification and vacuum stability against charge- and color-breaking minima; the direct
experimental limits, including dimension-five and dimension-six operators for proton decay, the measured
Higgs boson mass, and the LEP chargino bound; dark matter (DM), including the thermal relic density
and direct detection limits on the lightest supersymmetric particle (LSP); and the
flavor and precision observables, such as the $\mu\to e\gamma$ bound, the muon anomalous magnetic moment
$(g-2)_\mu$, and electric dipole moments (EDMs).  In this work, we take the simplest and most predictive
soft spectrum: a universal, constrained-MSSM-like (CMSSM) pattern in which a single
soft scalar mass $\mzero$ governs all squarks and sleptons, with no inter-generation
hierarchy.

Taken together, these constraints leave only a narrow, low-to-moderate-$\tb$ region of the
mini-split type. It is bounded from below by the proton lifetime bound, which forces the scalars heavy,
and from above by the observed Higgs mass. The surviving region demands a heavy universal scalar
mass and a bounded $\tb$. Within it, the lightest neutralino is a bino whose thermal relic over-closes the
Universe, so the universal model does not naturally provide a viable thermal neutralino DM candidate over most of the surviving parameter space, and DM
enters as a model-dependent input. Only in the minimal non-universal-Higgs-mass (NUHM) extension, where
$\mu$ is freed, does a light neutralino relic reappear, and present direct detection data then
exclude the well-tempered strip and leave only a $\sim1.1$~TeV higgsino.

In summary, we work with the effective $\rep{10}\oplus\overline{\rep{126}}$ Yukawa sector: the charged-fermion fit, the seesaw, and the dimension-five proton decay operators all follow from the fitted Yukawa couplings, while the colored Higgs sector is reduced to a single effective triplet scale $\MHC$ that sets the proton decay normalization and is pinned by gauge unification. A unique proton lifetime in a fully renormalizable completion would instead require the complete color triplet mass matrix together with the accompanying doublet-mixing factors, which lie beyond the effective treatment adopted here.

The paper is organized as follows. \Cref{sec:model} sets out the
model, including its Lagrangian, the $\rep{10}\oplus\overline{\rep{126}}$ Yukawa sector, the
universal soft terms, and our notation. \Cref{sec:theory} covers the theoretical
consistency constraints and \Cref{sec:walls} the direct experimental limits,
notably dimension-five and dimension-six proton decay operators. \Cref{sec:flavor} treats the flavor and precision observables. \Cref{sec:scan} presents the electroweakino sector in the free-$\mu$ and constrained-$\mu$ schemes and the resulting allowed parameter space, and \Cref{sec:dm} the fate of the lightest neutralino as DM in each scheme.
\Cref{sec:wedge} interprets the surviving region, compares it with prior
work, and lists its near-future signatures. \Cref{sec:concl} concludes with our findings.

\section{The model}\label{sec:model}

This section fixes the model and the notation: the
field content and Yukawa sector of the minimal supersymmetric $\SO{10}$, the
color triplet and doublet--triplet sector that mediates
proton decay and is tied to gauge coupling unification, the
universal (CMSSM-like) soft supersymmetry-breaking sector that is our
working hypothesis, and the inputs to the analysis.

\subsection{Field content and the Yukawa sector}\label{sec:model_field}

The matter sector of $\SO{10}$ is maximally economical: a single irreducible
spinorial $\rep{16}$ chiral superfield per generation accommodates all fifteen
SM Weyl fermions of that generation, along with a right-handed
neutrino $N^c$. We write the three generations as $\rep{16}_a$
with a family index $a=1,2,3$ acting on the generation space. Because a single multiplet
carries all the matter, the entire flavor structure of the theory is encoded in how the
three $\rep{16}_a$ couple to the Higgs sector.

In the \emph{minimal} model, the Yukawa couplings of the matter spinors arise from just
two Higgs representations, a $\rep{10}$ and a $\overline{\rep{126}}$,
through the renormalizable superpotential~\cite{Babu:1992ia}:
\begin{equation}
\label{eq:WY}
W_Y 
= 
\rep{16}_a\,\Bigl[ \,(Y_{10})_{ab}\,\rep{10}
+ (Y_{126})_{ab}\,\overline{\rep{126}} \, \Bigr]\,\rep{16}_b 
~,
\end{equation}
where the two Yukawa matrices $Y_{10}$ and $Y_{126}$ act on the family indices
$a,b=1,2,3$. The group theory of $\SO{10}$ forces both matrices to be
\emph{symmetric}~\footnote{The
product $\rep{16}\otimes\rep{16}=\rep{10}\oplus\rep{120}\oplus\overline{\rep{126}}$
contains the $\rep{10}$ and the $\overline{\rep{126}}$ in its symmetric part, while the antisymmetric $\rep{120}$ is absent in the minimal model.}, which is the
source of the model's predictive power.

The single pair of symmetric matrices $(Y_{10},Y_{126})$ therefore controls three
distinct sectors at once.
When the $\rep{10}$ and $\overline{\rep{126}}$ acquire electroweak-scale VEVs along
their SM doublet directions, \cref{eq:WY} produces the up-type, down-type,
charged-lepton, and Dirac-neutrino mass matrices as fixed linear combinations of the
same $Y_{10}$ and $Y_{126}$. Explicitly, writing $v^{10}_{u,d}$ and $v^{126}_{u,d}$ for
the up- and down-type Higgs doublet VEVs residing in the $\rep{10}$ and the $\overline{\rep{126}}$, we have
\begin{align}
\label{eq:massrel}
M_u &= Y_{10}\,v^{10}_u + Y_{126}\,v^{126}_u ~, \\
M_{\nu_{\rm D}} &= Y_{10}\,v^{10}_u - 3\,Y_{126}\,v^{126}_u ~, \\
M_d &= Y_{10}\,v^{10}_d + Y_{126}\,v^{126}_d ~, \\
M_e &= Y_{10}\,v^{10}_d - 3\,Y_{126}\,v^{126}_d ~,
\label{eq:massrel2}
\end{align}
where the factor $-3$ multiplying the $Y_{126}$ contribution to the leptonic masses is
the $SU(4)_c$ Clebsch--Gordan coefficient by which the $\overline{\rep{126}}$
distinguishes leptons from quarks (leptons carry $B-L=-1$, three times the quark value in
magnitude), and is the origin of the $\SO{10}$ analogue of the Georgi--Jarlskog
relations~\cite{Babu:1992ia,Fukuyama:2004pb}. The two symmetric matrices $Y_{10},Y_{126}$
thus appear in all four charged- and neutral-fermion mass matrices at once, which is what
makes the fit overdetermined. Fitting these two matrices to the measured
charged-fermion masses and the CKM matrix therefore leaves essentially no freedom (up to residual fit ambiguities~\footnote{These include discrete sign and phase branches, and the split of each fermion mass between the $\rep{10}$ and $\overline{\rep{126}}$, which leave the charged-fermion spectrum unchanged but are fixed, for the proton decay phase, by the color triplet sector below}). The same matrices thus control, through the unified
$\SO{10}$ structure, the flavor structure of the $B$-violating dimension-five
Wilson coefficients $C_L$ (for the operator $QQQL$) and $C_R$ (for $u^cu^cd^ce^c$) that
drive proton decay~\cite{Weinberg:1981wj,Sakai:1981pk,Dimopoulos:1981dw}; and, through the
large $\SO{10}$-breaking VEV of the $\overline{\rep{126}}$, the heavy Majorana mass
matrix of the right-handed neutrinos. The flavor structure that reproduces the quark
and lepton masses also fixes the proton decay amplitude, and the decay rate is
a prediction of the fitted theory rather than an adjustable parameter.

The right-handed-neutrino Majorana mass deserves separate mention. The
$\overline{\rep{126}}$ contains a SM singlet whose VEV
$v_R\equiv\langle\overline{\rep{126}}\rangle$ breaks $B\!-\!L$ at a high scale and, with
$M_{\nu_{\rm D}}$ the Dirac neutrino mass of \cref{eq:massrel}, realizes the type-I seesaw,
\begin{equation}
\label{eq:seesaw}
M_R = Y_{126}\,v_R 
~, \qquad
m_\nu \;\simeq\; -\,M_{\nu_{\rm D}}^{\mathsf T}\,M_R^{-1}\,M_{\nu_{\rm D}} 
~,
\end{equation}
a Majorana mass matrix for the right-handed neutrinos $N^c$ together with the light
neutrino mass matrix it induces.  The same $\overline{\rep{126}}$ VEV that feeds proton decay through the GUT-scale dynamics is therefore also responsible for the small neutrino masses.

We obtain the dimension-five Wilson coefficients $C_L,C_R$ and the realized colored Higgs phase $\phi_{13}$ (the relevant $CP$ phase of the color triplet coupling that enters the dimension-five amplitude) by fitting $Y_{10}$ and $Y_{126}$ to the charged-fermion masses and the CKM matrix at $\MGUT$ and contracting the fitted matrices through the doublet--triplet vacuum. Crucially, and in contrast to analyses that leave the colored Higgs phase free and scan over it, we \emph{derive} the phase in the effective fit, under the phase convention fixed below: the measured third-generation $b$ and $\tau$ Yukawas are real, which forces $(Y_{10})_{33}$ and $(Y_{126})_{33}$ to be real, so their relative sign is a discrete $\pi$ flip rather than a tunable $CP$ phase and $\phi_{13} \approx 0$.  Varying the residual fit ambiguities still leaves $\phi_{13} \approx 0$, a robust prediction of the model.

More concretely, the third-generation down-quark and charged-lepton masses fix the relevant couplings through the $SU(4)_c$ relations $m_b\propto(Y_{10})_{33}+(Y_{126})_{33}$ and $m_\tau\propto(Y_{10})_{33}-3\,(Y_{126})_{33}$. So two real measured masses fix two real couplings, $(Y_{10})_{33},(Y_{126})_{33}\in\mathbb{R}$.  The color triplet ratio inherits this reality, $C_R/C_L=1+\mathcal{O}\!\big((Y_{126}/Y_{10})_{33}\big)\in\mathbb{R}$ with $(Y_{126}/Y_{10})_{33}\simeq0.06$, so $\phi_{13}=\arg(C_R/C_L)$ is fixed to the discrete values $0$ or $\pi$, not a continuous phase, the realized signs giving $\phi_{13}\approx0$. The color triplet couplings also carry mixing-matrix phases that the real diagonal third-generation masses do not by themselves fix.  An explicit matrix-level fit of $Y_{10}$ and $Y_{126}$ including the triplet mixing nonetheless confirms $\phi_{13}\approx0$: at the GUT-scale benchmark, the third-generation entries come out $(Y_{10})_{33}\approx0.060$ and $(Y_{126})_{33}\approx-0.0038$ (both real, with $\arg[(Y_{126}/Y_{10})_{33}]=\pi$ a discrete sign flip and $|(Y_{126}/Y_{10})_{33}|\approx0.06$), giving $|C_R/C_L| \approx 1.07$ and $\phi_{13}=\arg(C_R/C_L) \approx 0$. To go beyond the third-generation argument, we fit $Y_{10}$ and $Y_{126}$ to the charged fermion masses together with the complete CKM matrix, its magnitudes and its $CP$ phase, and evaluate $\arg(C_R/C_L)$ over the residual fit ambiguities.  The realized phase has a central branch sharply clustered near zero, with $96\%$ of the fits within $10^\circ$, together with a small $\sim4\%$ population on the discrete $\pi$ branch; the overall circular spread (standard deviation) is $0.40$~rad, dominated by that $\pi$-branch tail, with central value $\phi_{13}=-0.003$~rad and $|C_R/C_L|\approx1.1$.  The CKM phase enters $C_R/C_L$ only through off-diagonal terms weighted by $|(Y_{126}/Y_{10})_{33}|\approx0.06$ and the small quark mixing, so it broadens the distribution but does not move its center off the real axis.  The prediction $\phi_{13}\approx0$ is therefore a property of the full flavor structure, not of the third generation alone.

\subsection{The color triplet sector and unification}\label{sec:model_triplet}

The light Higgs doublets of the MSSM sit inside the $\rep{10}$ (and, through mixing,
the $\overline{\rep{126}}$). Their unavoidable companions are the color triplet,
weak-singlet Higgs states and, due to supersymmetry, the corresponding colored higgsino
of mass $\MHC$. It is the \emph{exchange} of this colored higgsino between two
quark--squark legs that generates the $B$-violating dimension-five operators,
encoded in the effective superpotential~\cite{Weinberg:1981wj,Sakai:1981pk,Dimopoulos:1981dw},
\begin{equation}
\label{eq:dim5}
\begin{split}
W_5
={}&
\frac{1}{\MHC}\Bigg[\,\frac12\,C_L^{ijk\ell}\,(Q_i Q_j)(Q_k L_\ell)
+ C_R^{ijk\ell}\,u^c_i\,e^c_j\,u^c_k\,d^c_\ell\,\Bigg]
 ~,
\end{split}
\end{equation}
where $i,j,k,\ell$ are the generation indices and the Wilson coefficients
$C_L\propto Y_{10,126}^2$ and $C_R\propto Y_{10,126}^2$ are fixed in both their overall
size and their flavor structure by the same fitted Yukawa matrices, divided by the
colored higgsino mass $\MHC$. Being dimension-five rather than dimension-six, these are
suppressed by only a single power of the heavy mass, $\sim1/\MHC$, rather than two, which
is why supersymmetric proton decay tends to be too fast. The dressed decay rate scales as
$\Gamma\propto\MHC^{-2}$, making $\MHC$ the single most important GUT-scale input to the
proton lifetime to be examined in \cref{sec:walls}.

In a fully specified GUT Higgs sector, $\MHC$ is a particular combination of the
$\rep{210}/\rep{126}$ VEVs and superpotential couplings that break $\SO{10}$, and the
doublet--triplet splitting problem is the requirement that the doublets remain light
while the triplets are heavy. Rather than commit to one such sector, we fix $\MHC$ from
gauge-coupling unification: in the supersymmetric $\SO{10}\to$ MSSM desert, the three
SM gauge couplings run from their measured low-scale values, meet at the
unification scale $\MGUT\approx1.4\times10^{16}$~GeV with a unified coupling
$\alpha_{\rm GUT}\approx1/24$~\cite{Georgi:1974yf,Dimopoulos:1981zb,Sakai:1981gr,Dimopoulos:1981yj}, and
the color triplet threshold modifies the running through the specific combination of
$3\alpha_2^{-1}-2\alpha_3^{-1}-\alpha_1^{-1}$~\cite{Hisano:1992jj} that isolates the triplet's
contribution alone. Demanding that the couplings still unify, given the precisely
measured $\alpha_s(M_Z)$, confines $\MHC$ to a two-sided band rather than leaving it free.
Because the same $\MHC$ sets the proton decay rate ($\Gamma\propto\MHC^{-2}$), unification and
proton decay are correlated.  We construct the band, and work out how it interlocks with
the proton decay bound, in \cref{sec:theory}. Though in a complete Higgs
sector $\MHC$ can exceed $\MGUT$, we impose $\MHC\le\MGUT$ as a calculational convention, 
treating the scale at which the couplings meet as the
natural upper limit for any state in the unified multiplets.

\subsection{The universal soft supersymmetry-breaking sector}\label{sec:model_soft}

Supersymmetry must be broken, and the breaking is parametrized by the soft Lagrangian
\begin{equation}
\label{eq:Lsoft}
\begin{split}
-\,\mathcal{L}_{\rm soft}
={}&
\sum_i m^2_i\,\bigl|\tilde\phi_i\bigr|^2
+ \left(\frac{1}{2}\sum_{a=1}^{3} M_a\,\lambda_a\lambda_a
+ \sum_f A_f\,\tilde{\overline f}\,\tilde f\,H + \text{H.c.}\right)
~,
\end{split}
\end{equation}
collecting the scalar mass-squareds $m^2_i$ of the squarks, sleptons and Higgs scalars
$\tilde\phi_i$; the Majorana masses $M_a$ of the bino, wino and gluino $\lambda_a$; and
the trilinear scalar couplings $A_f$. We adopt the simplest and most predictive pattern,
the \emph{universal} or CMSSM ansatz, which fixes these soft terms at
the GUT scale to
\begin{equation}
\label{eq:cmssm_bc}
m^2_{\tilde f} = \mzero^2\,\mathbbm{1}
~, \qquad
M_1 = M_2 = M_3 = \mhalf
~, \qquad
A_f = A_0\,Y_f
~,
\end{equation}
with the universal scalar mass $\mzero$, the universal gaugino mass $\mhalf$, and the
universal trilinear $A_0$. The scalar masses, the gaugino mass, and $A_0$ in \cref{eq:cmssm_bc} are the boundary values imposed at $\MGUT$.  The low-scale spectrum follows by renormalization-group running, and $|\mu|$ (and the soft parameter $B$) are the low-scale outputs fixed by electroweak symmetry breaking (EWSB), not inputs. Here, $A_f$ is the trilinear soft coupling of the
sfermion $\tilde f$, $Y_f$ the corresponding SM fermion Yukawa coupling, and
the boundary condition fixes $A_t(\MGUT)=A_0$ (and likewise for the other flavors).
Here $\mzero^2\,\mathbbm{1}$ is proportional to the identity
in flavor space, so the squark and slepton soft masses are flavor-blind: there is no
inter-generation hierarchy and no new flavor structure beyond the Yukawa (CKM-induced)
one. This is what ``universal'' means, and, as noted in \cref{sec:intro}, it is the
most predictive soft sector because it leaves no room to weaken proton decay by
decoupling certain generations, yielding a clean lower bound on $\mzero$. Two
further parameters complete the specification: $\tb \equiv v_u/v_d$, the ratio of the up- and
down-type Higgs VEVs; and the sign of the supersymmetric Higgs mass
parameter $\mu$, since under the CMSSM boundary conditions $|\mu|$ is not free but
fixed by the radiative EWSB minimization condition.

Three low-scale consequences of \cref{eq:cmssm_bc} matter for what follows. First, the
common gaugino mass runs to the low-scale ratio
\begin{equation}
\label{eq:gaugino_ratio}
M_1 : M_2 : M_3 \simeq 1 : 2 : 6 
~,
\end{equation}
fixed by the SM gauge $\beta$-functions, where $M_1$, $M_2$, and $M_3$ are
the bino, wino, and gluino soft masses. Both the wino and the gluino that dress the
proton decay loop are therefore tied to the single scale $\mhalf$. Second, the universal
trilinear sets the stop mixing
\begin{equation}
\label{eq:Xt}
\Xt 
=
A_t - \mu\cot\beta 
~,
\end{equation}
where $A_t$ is the low-scale stop trilinear ($A_0$ being its GUT-scale input), and $\Xt$ acts
in the $2\times2$ stop mass-squared matrix. The parameter $\Xt$ drives the radiative correction to the
Higgs mass and is essential to reaching $m_h=125$~GeV to be detailed in \cref{sec:walls}. Third,
because the scalars are free to be heavy while the gauginos and higgsinos are kept near
the TeV scale, the natural regime once the proton bound pushes $\mzero$ up is a
\emph{mini-split} spectrum~\cite{Arkani-Hamed:2012fhg,Arvanitaki:2012ps}: one in which the scalars are far heavier than the
electroweakinos, which both dress the proton loop and preserve unification.

For the phenomenological scan we treat $\{\mzero,\,M_2,\,\mu,\,\tb,\,\Xt\}$, together
with $\MHC$, as the working variables, with $M_1$ and $M_3$ following from \cref{eq:gaugino_ratio}.
We sample $M_2$ and $\mu$ directly as effective low-scale inputs and impose only the
sign requirement $\mu>0$ for radiative EWSB. Solving the EWSB conditions point by point, under which $M_2$ and $|\mu|$ are determined by $\{\mzero,\mhalf,A_0,\tb\}$ rather than chosen
independently, is imposed separately in the constrained-$\mu$ cross-check in \cref{sec:scan}.

\subsection{Inputs, benchmarks, and notation}\label{sec:model_inputs}

\Cref{tab:inputs} collects the inputs to the analysis: the scanned soft-breaking
parameters with their ranges (top block) and the fixed SM and hadronic
inputs that enter the proton decay rate and the Higgs mass (bottom block). The scanned
parameters are exactly the working set of \cref{sec:model_soft}.

\begin{table*}[htbp]
\centering
\caption{Inputs to the analysis. Top block: the scanned soft-breaking parameters and
their ranges (sampled uniformly in $\log_{10}$, except $\tb$, which is linear), with $M_1,M_3$ fixed by the GUT
ratio \cref{eq:gaugino_ratio}. Bottom block: the fixed SM and hadronic
inputs, each with its source. For the observed Higgs mass $m_h$, a wider acceptance window is used instead in the scan, reflecting the \emph{theory} uncertainty of the resummed $m_h$, not an experimental error.}
\label{tab:inputs}
\bigskip
\begin{tabular}{@{}l l l l@{}}
\toprule
Symbol & Value / range & Unit & Source / remark \\
\midrule
\multicolumn{4}{@{}l}{\emph{Scanned soft-breaking inputs}} \\
$\mzero$ & $0.5$--$3000$ & TeV & universal scalar mass \\
$M_2$    & $0.3$--$5$    & TeV & wino mass; $M_1{:}M_2{:}M_3\simeq1{:}2{:}6$ \\
$|\mu|$  & $0.3$--$5$    & TeV & effective input; sign $\mu>0$ imposed \\
$\tb$    & $2$--$50$     & ---  & ratio of Higgs VEVs, $v_u/v_d$ \\
$\Xt$    & $|\Xt|\le\min(\text{pos.},\,6\,\mzero)$ & --- & stop mixing $=A_t-\mu\cot\beta$; \\
& & & imposing stop-positivity bound and \\
& & & charge- and color-breaking stability imposed \\
$\MHC$   & up to $\MGUT$ & GeV & colored higgsino mass; pinned by unification \\
\midrule
\multicolumn{4}{@{}l}{\emph{Fixed SM and hadronic inputs}} \\
$\alpha_s(M_Z)$ & $0.1180\pm0.0009$ & --- & PDG 2026~\cite{PDG26} \\
$m_t$           & $172.6\pm0.27$    & GeV & PDG 2026~\cite{PDG26} \\
$m_h$           & $125.13\pm0.11$   & GeV & PDG 2026~\cite{PDG26} (listing average) \\
$\beta_H$       & $0.0144$          & GeV$^3$ & lattice~\cite{Aoki:2017puj} (proton decay matrix element) \\
$f_\pi^{(0)}$   & $86.8$            & MeV & lattice~\cite{Aoki:2017puj} \\
\bottomrule
\end{tabular}
\end{table*}

We treat the electroweakino sector in two ways: (i) the constrained-$\mu$ scheme, the prediction of the universal hypothesis, in which $|\mu|$ and the spectrum are determined point by point from the GUT-scale boundary
conditions \cref{eq:cmssm_bc} by renormalization-group (RG) running plus a
tree-level EWSB treatment (\cref{sec:scan_strict}), and (ii) a free-$\mu$ scan, a minimal
NUHM extension of the previous one, in which $M_2$ and $|\mu|$ are sampled directly as
low-scale inputs (with $\mu>0$ standing in for radiative EWSB) so that a light electroweakino can
be probed (\cref{sec:scan_headline}). Two ingredients of the calculation are
approximate. The Higgs boson mass is evaluated with an effective field theory-resummed (EFT-resummed) treatment, which uses a one-loop leading-log expression calibrated point-by-point to \texttt{FeynHiggs~2.19}~\cite{Heinemeyer:1998yj}, which lowers it by $\sim6$--$9$~GeV for stop masses of tens of TeV (where the bare leading-log
overshoots).  The proton decay rate is normalized following the
benchmark calculation of \cite{Goto:1998qg}. The dimension-five $p\to K^+\bar\nu$ rate for the minimal $\SO{10}$ color triplet sector is not available from a public code.  The amplitude is evaluated from the fitted Yukawa couplings, contracted through the doublet--triplet vacuum, and normalized to the published Goto--Nihei rate, as detailed in \cref{sec:walls_dim5}. The dimension-six gauge-boson decay mode is unobservably slow at these scales and is not imposed. The individual constraints are
developed in \cref{sec:theory,sec:walls,sec:flavor}, the electroweakino sector in \cref{sec:scan}, and DM in
\cref{sec:dm}.

\section{Gauge unification and vacuum stability}\label{sec:theory}

The constraints in this section are theoretical consistency
conditions, required for the universal minimal supersymmetric $\SO{10}$ spectrum to be physical. A point in the parameter space that violates any of them describes a
tachyonic stop, a charge- and color-breaking (CCB) vacuum, or a spectrum whose gauge couplings do
not unify, and is thus removed.
Gauge coupling unification plays a second role as well: it fixes the
color triplet (colored higgsino) mass band $\MHC$ that enters the proton decay rate,
linking this section to the direct experimental limits of \cref{sec:walls}.

\subsection{Gauge-coupling unification and the \texorpdfstring{$\MHC$}{MHC} band}\label{sec:theory_unif}

In the supersymmetric $\SO{10}\to$ MSSM desert, the three SM gauge couplings must, given
their precisely measured low-scale values, converge to a single value at the unification scale
$\MGUT\approx1.4\times10^{16}$~GeV~\cite{Georgi:1974yf,Dimopoulos:1981zb,Sakai:1981gr,Dimopoulos:1981yj}. Here
$\MGUT$ is not an input but an output, the scale at which $\alpha_1$ and $\alpha_2$ are computed
to cross under the supersymmetric running of the measured electroweak couplings.  With only a weak
dependence on the sparticle scale, it comes out at $1.427\times10^{16}$~GeV for $\mzero=12$~TeV, and varies by only $1.5\%$ over the whole surviving region $\mzero=10$--$30$~TeV.  This is the value used throughout, in the cap $\MHC\le\MGUT$ of \cref{eq:mhc_band} and in the dimension-six rate of \cref{sec:walls_dim6} alike. Writing
$\alpha_i^{-1}(\mu)$ for the inverse couplings of $U(1)_Y$, $SU(2)_L$ and $SU(3)_c$ ($i=1,2,3$) in
the standard grand unified normalization, the color triplet threshold modifies the running through
the specific linear combination~\cite{Hisano:1992jj}
\begin{equation}
\label{eq:unif_comb}
3\,\alpha_2^{-1} - 2\,\alpha_3^{-1} - \alpha_1^{-1} 
~,
\end{equation}
which is constructed to be blind to the heavy GUT-breaking gauge bosons and adjoints and to isolate
exactly the contribution of the color triplet, weak-singlet Higgs partner.  Ref.~\cite{Hisano:1992jj} gives the systematic treatment of the dimension-five proton decay dressing together with the color triplet threshold matching used throughout this study.  Because \cref{eq:unif_comb} retains the triplet but cancels the rest of the GUT
spectrum, its measured value translates the requirement of unification directly into a condition on
the triplet mass. The blindness to $\MGUT$ is twofold. The coefficients in \cref{eq:unif_comb} sum to
zero, so the unified coupling and every \emph{complete} GUT multiplet, including the superheavy gauge
bosons whose mass defines $\MGUT$, contribute equally to the three $\alpha_i^{-1}$ and cancel;
and those same coefficients annihilate the slope of the running between $\MHC$ and $\MGUT$ (where the
spectrum is the MSSM plus the color triplet), which removes the $\ln\MGUT$ term. What survives is set
by the color triplet alone, an \emph{incomplete} multiplet split from its light doublet partner.
Consequently, $\MHC$, and with it the proton decay bound, is independent of $\MGUT$ and of the
GUT-scale thresholds.  Instead, $\MGUT$ enters only the subdominant dimension-six mode
(to be discussed in \cref{sec:walls}) and the condition $\MHC\le\MGUT$ below.

The practical consequence is sharp.  Demanding that the three couplings still meet, given the
accurately determined strong coupling $\alpha_s(M_Z)$ whose LEP-era determination first sharpened
supersymmetric unification into a quantitative success~\cite{Amaldi:1991cn,Langacker:1991an,Ellis:1990wk},
fixes the colored higgsino mass $\MHC$ to a two-sided band rather than leaving it free. Our
working window is
\begin{equation}
\label{eq:mhc_band}
\MHC \in [\,0.58,\,1.43\,]\times10^{16}~\text{GeV}
~,
\end{equation}
a factor-$\sim2$ band fixed by a two-loop running of \cref{eq:unif_comb} following the threshold
analysis of \cite{Hisano:1992jj}. The lower edge is the value below which the couplings overshoot before meeting, the upper edge the value above which they undershoot. The width of this band comes from the measured strong coupling. Propagating $\alpha_s(M_Z)=0.1180\pm0.0009$~\cite{PDG26} through the unification condition gives $\MHC\in[0.70,1.35]\times10^{16}$~GeV, a factor $1.9$, at the reference scalar mass $\mzero\simeq12$~TeV near the proton decay bound. Each edge is then widened by a factor $1.22$, a conservative allowance for the two-loop and GUT threshold corrections of the size found in Ref.~\cite{Hisano:1992jj}, and the upper edge is capped at $\MGUT$; this gives the quoted \cref{eq:mhc_band}, a factor $2.5$. The band narrows as $\mzero$ rises, since the upper edge saturates the cap while the lower edge continues to rise. Because $\tau_p\propto\MHC^{2}$ and $\mzero\propto\tau_p^{1/4}\propto\MHC^{1/2}$, it maps to a factor $\sqrt{1.43/0.58} \simeq 1.6$ on the proton decay lower bound on $\mzero$, collected with the other uncertainties in \cref{sec:walls_dim5}. That bound is evaluated at the most favorable in-band value $\MHC=\min(\MHC^{\rm hi},\MGUT)$, where $\MHC^{\rm hi}$ is the upper edge of \cref{eq:mhc_band}, which gives the weakest bound, so any smaller $\MHC$ within the band would only strengthen it.  A complete, model-dependent GUT-threshold calculation would require the full colored Higgs sector and lies beyond the effective single-$\MHC$ treatment adopted here.

This band is the most consequential entry in this section. The same $\MHC$ that fixes
unification also sets the proton decay rate, $\tau_p\propto\MHC^2$.  So the obvious way to
lengthen the proton lifetime by raising $\MHC$ is strongly disfavored: a larger $\MHC$ spoils
unification because the couplings then fail to meet. The normalization of the
proton decay bound is therefore fixed by unification rather than left
adjustable, a tie made quantitative in \cref{sec:walls}.

We additionally impose an upper bound on the colored higgsino mass at the unification scale, $\MHC\le\MGUT$. In a fully specified GUT-Higgs sector, however, the triplet mass is a particular combination of VEVs and superpotential couplings and can in principle exceed $\MGUT$.  But in the absence of such a detailed model, we treat $\MGUT$ as a natural upper limit for any state living in the unified multiplets. This bound ties the upper edge of the band to the same running that defines $\MGUT$. Throughout this work, $\MHC$ denotes an effective single scale standing for the color triplet combination that controls both unification and the dimension-five rate.  In a complete GUT-Higgs sector, these are entries of the inverse color triplet mass matrix together with doublet--triplet mixing.  The single-$\MHC$ treatment is the leading parametrization, and resolving the full matrix structure is beyond the scope of the present analysis.

To estimate the effect of this assumption, take the simplest extension beyond a single triplet: two color triplet pairs with masses $M_1\le M_2$ and spread $r\equiv M_2/M_1$.  Gauge unification constrains the sum of their logarithms, so it fixes their geometric mean, $\MHC^{\rm unif}=(M_1 M_2)^{1/2}$, whereas the dimension-five amplitude runs through the inverse mass matrix, $1/\MHC^{\rm dec}=|w_1^2/M_1\pm w_2^2/M_2|$ with $w_1^2+w_2^2=1$, the weights with which the operator couples to the two triplets.  For comparable couplings and no cancellation, the decay scale is the harmonic mean, and the two effective scales differ by
\begin{equation}
\label{eq:mhc_decouple}
\frac{\MHC^{\rm unif}}{\MHC^{\rm dec}}=\frac{1+r}{2\sqrt r}=\cosh\!\Big(\tfrac12\ln r\Big)
~,
\end{equation}
a factor of $1.15$, $1.7$, and $2.8$ for $r=3$, $10$, and $30$.  Since $\mzero\propto\MHC^{1/2}$, such a decoupling shifts the proton decay lower bound on $\mzero$ by at most $(\MHC^{\rm unif}/\MHC^{\rm dec})^{1/2}\approx1.1$--$1.7$ over the same range.  In the generic case, the decay scale is the \emph{smaller} of the two (the harmonic mean lies below the geometric mean), which would \emph{strengthen} the bound.  The one loophole is a relative sign between the two terms, $1/\MHC^{\rm dec}=|w_1^2/M_1-w_2^2/M_2|$: a tuned near-cancellation can raise $\MHC^{\rm dec}$ and suppress the rate, the known freedom by which an extended color triplet sector can evade proton decay.  We do not exploit this tuning in this work.  The single-$\MHC$ choice is an effective parametrization of the minimal, most predictive scenario, not a general property of every supersymmetric $\SO{10}$ model, and a more general sector would multiply the rate by a model-dependent factor that we do not scan.

\subsection{Vacuum stability: stop positivity and the CCB bound}\label{sec:theory_ccb}

The stop trilinear coupling $\Xt=A_t-\mu\cot\beta$ controls the top-squark left--right mixing and
enters the one-loop Higgs mass through the polynomial $f(x)=x^2\bigl(1-x^2/12\bigr)$ with
$x=\Xt/M_S$, where $M_S$ is the stop mass scale and here taken to be $\mzero$. The function $f(x)$ peaks at the maximal-mixing
point $|\Xt|=\sqrt6\,M_S \approx 2.449 M_S$, with $f(\sqrt6)=3$ and turns \emph{negative} for
$|\Xt|>\sqrt{12}\,M_S\approx3.46\,M_S$.  The sign of $f$ is what matters physically: for $|\Xt|/M_S<\sqrt{12}$ the mixing \emph{raises} $m_h$ (most strongly at $\sqrt6$), while in the negative-$f$ window $|\Xt|/M_S>\sqrt{12}$ it \emph{lowers} $m_h$. Reaching $m_h=125$~GeV needs the positive-$f$ region, so the negative-$f$ window must be excluded; otherwise, an arbitrarily heavy $\mzero$ could be tuned back to $125$~GeV by over-large mixing, spuriously erasing the upper bound on $\mzero$. The restriction on the mixing comes instead from the following two physical criteria.

\emph{Stop positivity} requires both eigenvalues of the full $2\times2$ top-squark mass-squared
matrix to be positive (no tachyonic stop). In the universal scenario, with both soft stop masses equal
to $\mzero$ and the electroweak $D$-terms negligible against $\mzero^2$, the smaller eigenvalue is $\approx\mzero^2+m_t^2-m_t|\Xt|$ in the off-diagonal-dominant approximation. 
So the positivity reads $|\Xt|\lesssim(\mzero^2+m_t^2)/m_t$. This
is essentially non-binding: because the stops are heavy here ($\mzero\gg m_t$), they stay
non-tachyonic even under enormous mixing, the bound being far above the
$\sqrt{12}\approx3.46$ where the Higgs physics matters (e.g., $|\Xt|/\mzero\lesssim\mzero/m_t\sim17$ at $\mzero\sim3$~TeV).

\emph{Charge- and color-breaking metastability} is the operative condition. Physically, the ordinary
electroweak vacuum must not be destabilized by a deeper minimum along a CCB
field direction: a large soft trilinear can tilt the scalar potential so that a vacuum where the
stop fields acquire VEVs, thereby breaking color and electric charge at once, lies below the standard one, which
would be catastrophic. The constraint on the soft trilinears was first developed by
Fr\`ere--Jones--Raby and Claudson--Hall--Hinchliffe~\cite{Frere:1983ag,Claudson:1983et}.  We impose it
in the empirical sufficient-stability form of Casas, Lleyda and Mu\~{n}oz~\cite{Casas:1995pd},
\begin{equation}
\label{eq:ccb}
\begin{gathered}
|A_t|^2 + 3\mu^2 \;<\; 3\bigl(m^2_{Q_3}+m^2_{U_3}+m^2_{H_u}\bigr)
~, \qquad
A_t=\Xt+\mu\cot\beta
~,
\end{gathered}
\end{equation}
evaluated in the universal scenario $m^2_{Q_3}=m^2_{U_3}=\mzero^2$.  Its
right-hand side still depends on the up-type Higgs soft mass $m^2_{H_u}$, which runs and is not
fixed at the stop scale.  So we bracket it: the primary cut uses $m^2_{H_u}=\mzero^2$, and we
also keep the conservative choice $m^2_{H_u}=0$, which drops that term and gives the
tightest bound. With $m^2_{H_u}=\mzero^2$, the constraint maximizes the mixing at
$|\Xt|/\mzero<3.06$; the conservative $m^2_{H_u}=0$ value would tighten this only to
$\sqrt6\approx2.45$.  Either upper bound sits below the negative-$f$ threshold
$|\Xt|/\mzero=\sqrt{12}\approx3.46$.  Therefore, CCB stability forbids the entire window where
the mixing would lower $m_h$.  In other words, it is the vacuum stability, not the stop positivity, that closes off that region.

\subsection{The \texorpdfstring{$\mu>0$}{mu>0} sign requirement}\label{sec:theory_mu}

We sample the higgsino mass parameter $\mu$ as an effective input and impose only
its \emph{sign}, $\mu>0$, as a conservative constraint. Such a sign choice, which does not solve the electroweak-minimization condition, reproduce $M_Z$, or test spectrum viability, is a weak condition, but it connects to a deeper issue. Under the universal (CMSSM) boundary conditions, $|\mu|$ is not a free input but is
determined by $\{\mzero,\mhalf,A_0,\tb\}$ through the electroweak minimization condition.
A more rigorous scan for radiative EWSB, and the question of whether the light-$\mu$ region is actually realized, is examined in the constrained-$\mu$ cross-check in \cref{sec:scan}.

\subsection{The consistency cuts in the parameter space}\label{sec:theory_figs}

\begin{figure}[htbp]
\centering
\includegraphics[width=0.8\columnwidth]{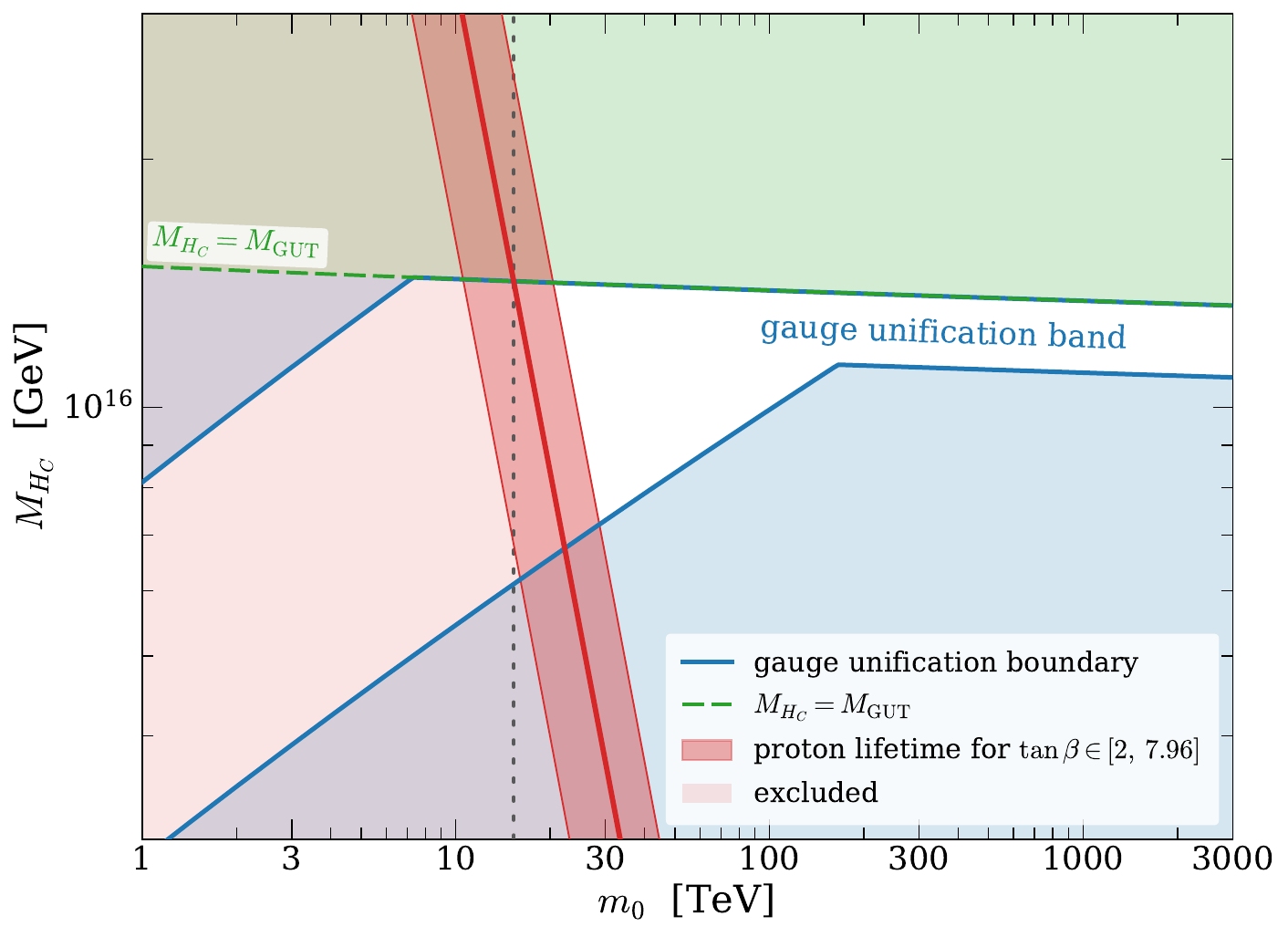}
\caption[Unification and proton decay in $(\mzero,\MHC)$.]{Gauge unification and proton decay constraints
together in the $(\mzero,\MHC)$ plane. Demanding that
the combination $3\alpha_2^{-1}-2\alpha_3^{-1}-\alpha_1^{-1}$ unify (given the measured $\alpha_s$)
fixes the colored higgsino mass to the band $\MHC\in[0.58,1.43]\times10^{16}$~GeV of
\cref{eq:mhc_band}, subject to the assumed bound $\MHC\le\MGUT$.  The proton lifetime lower bound rises with $\tb$ and is drawn in red for $\tb \in [2,8]$ (thick red at 5).  Here, we take $M_2=\mu=1$~TeV.  The colored regions are excluded: when $\MHC$ lies \emph{outside} the unification band (light blue), when $\MHC>\MGUT$ (light green), and when the proton lifetime $\tau_p<6.61\times10^{33}$~yr (light red).}
\label{fig:bnd_mhc}
\end{figure}

\begin{figure}[htbp]
\centering
\includegraphics[width=0.8\columnwidth]{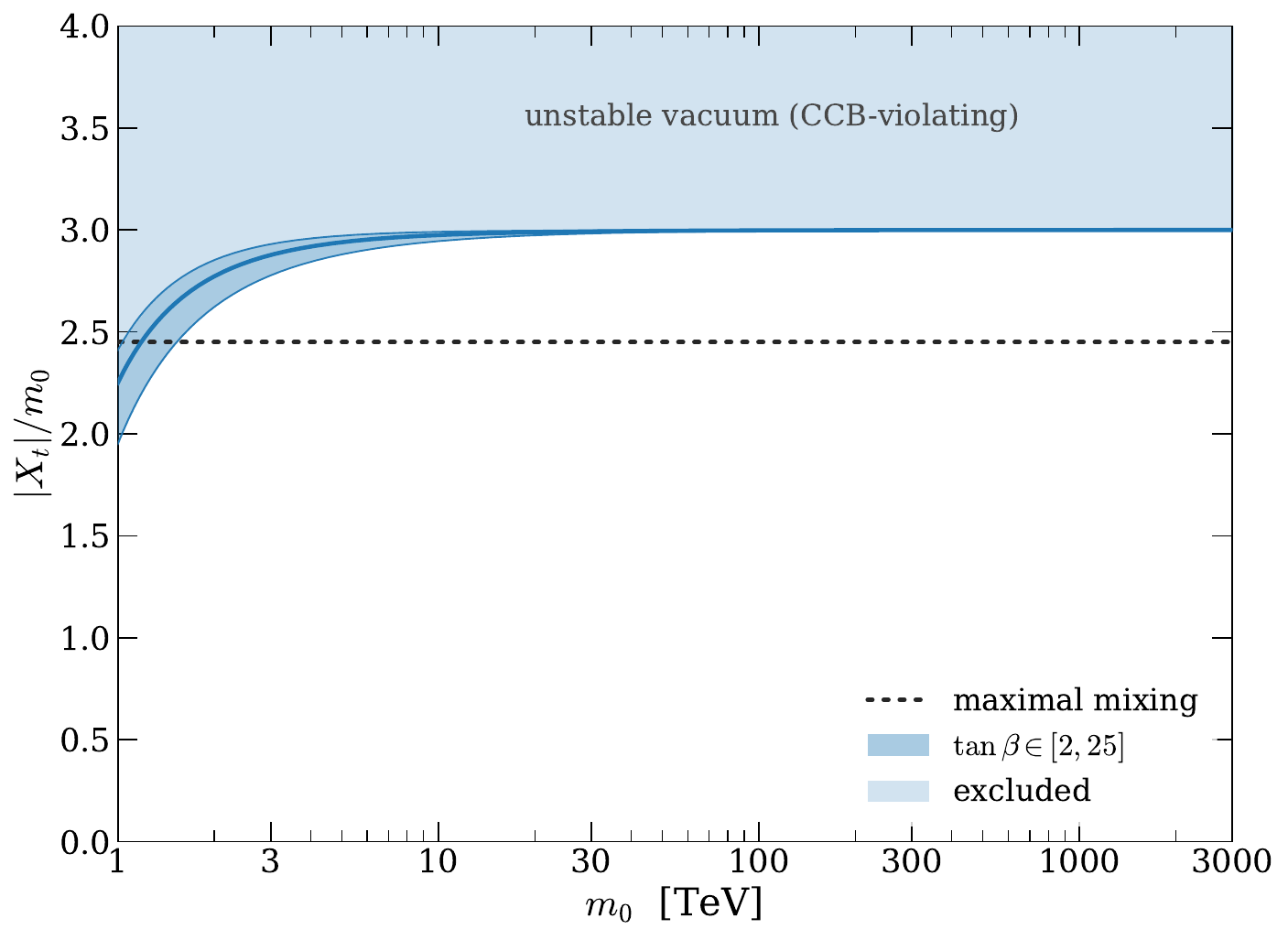}
\caption[Vacuum-stability consistency condition in isolation.]{The vacuum stability consistency
condition (stop positivity together with the Casas--Lleyda--Mu\~{n}oz CCB bound,
\cref{eq:ccb}). The dotted line marks the maximal-mixing reference value
$|\Xt|/\mzero=\sqrt6$, the stop mixing that maximizes the one-loop contribution to the
light-Higgs mass, shown for reference. The CCB bound of Casas, Lleyda and Mu\~{n}oz~\cite{Casas:1995pd} is given by the solid curve, sitting just above the dotted line and reaching the mixing at $|\Xt|/\mzero\simeq3$,
still below the negative-$f$ threshold of $\sqrt{12}\approx3.46$. The region above the CCB curve has an
unstable (CCB-violating) vacuum and is excluded. In this plot, we take $\tb \in [2,25]$ (thick blue at 5), $\mu=1$~TeV, and top mass $m_t=172.6$~GeV.}
\label{fig:con_ccb}
\end{figure}

\Cref{fig:bnd_mhc,fig:con_ccb} display the two consistency conditions in isolation, before any
experimental limit is overlaid.  Neither is a bound on $\mzero$ by itself: unification fixes the
$\MHC$ band (the empty region in \cref{fig:bnd_mhc}), which in turn will normalize the proton decay bound of
\cref{sec:walls}, while stop positivity together with CCB limits the stop mixing
(\cref{fig:con_ccb}), which controls where the upper edge of the Higgs mass constraint falls
rather than bounding $\mzero$ directly.

The lower edge of the unification band in \cref{fig:bnd_mhc} rises with $\mzero$ because heavier
scalars push the color triplet mass required by gauge unification upward toward the unification scale
(the mini-split spectrum sharpens the meeting of the couplings), until the band is forced to saturate the bound $\MHC\le\MGUT$ and falls with the slowly decreasing $\MGUT$.  The upper (lower) edge reaches that limit around $\mzero \sim 7$~TeV ($\sim 166$~TeV).

\section{Proton decay and the Higgs-mass bound}\label{sec:walls}

The direct experimental limits considered in this section
constrain the universal scalar mass $\mzero$ from both sides. Supersymmetric proton
decay sets a lower bound on $\mzero$ that rises with $\tb$; the measured $125$~GeV
Higgs mass sets an upper bound that falls with $\tb$. The two cross at
$\tb \simeq 8.0$, so the allowed region is confined to low-to-moderate $\tb$ and heavy scalars. The LEP
chargino limit constrains only the light electroweakino sector and does not bound $\mzero$.

\subsection{Dimension-five proton decay}\label{sec:walls_dim5}

In a supersymmetric GUT, the dominant $B$-violating interactions are not gauge-boson
exchange but a pair of dimension-five four-fermion operators~\footnote{These operators were
identified in supersymmetric grand unification in 1981--82 by Weinberg, Sakai and Yanagida,
and Dimopoulos, Raby and Wilczek~\cite{Weinberg:1981wj,Sakai:1981pk,Dimopoulos:1981dw}. Their
appearance makes the supersymmetric proton lifetime far shorter than the dimension-six
estimate, the recurrent ``dimension-five problem'' of supersymmetric GUTs.},
\begin{equation}
\label{eq:dim5ops}
  \mathcal{O}_L = (QQQL),\qquad \mathcal{O}_R = (u^c u^c d^c e^c)~,
\end{equation}
with $Q,L$ the quark and lepton $SU(2)_L$ doublet superfields and $u^c,d^c,e^c$ the conjugate
singlets. They arise from exchange of the colored higgsino---the color triplet, weak-singlet
partner of the Higgs doublets, of mass $\MHC$ fixed by gauge-coupling unification in
\cref{sec:theory}---and, being dimension five, are suppressed by a single power $1/\MHC$
rather than the $1/\MGUT^{2}$ of the gauge mode, which makes supersymmetric proton decay fast.

\paragraph{Wilson coefficients.}
Integrating the color triplet out of the superpotential generates the two operators with
seesaw-like Yukawa sandwiches~\cite{Sakai:1981pk} in the \emph{same} matrices
$Y_{10},Y_{126}$ that reproduce the charged-fermion masses and the CKM matrix. With the
$\rep{10}\oplus\overline{\rep{126}}$ Higgs fields the left-handed coefficient is~\cite{Fukuyama:2004pb}
\begin{equation}
\label{eq:CL}
  C_L = (Y_{10},Y_{126})\,\widetilde{\mathcal M}_C^{-1}\,(Y_{10},Y_{126})^{\mathsf T}~,
\end{equation}
with $\widetilde{\mathcal M}_C$ the color triplet mass matrix (lightest eigenvalue $\MHC$).
The right-handed coefficient shares this structure, with the left vertex dressed by the
triplet mixing,
\begin{equation}
\label{eq:CR}
\begin{gathered}
  C_R=\Big(Y_{10}-a\,Y_{126},\;\; b\,Y_{126}\Big)\,\widetilde{\mathcal M}_C^{-1}\,\binom{Y_{10}}{Y_{126}}~,
  \qquad
  a=\frac{(\mathcal M_C)_{13}}{(\mathcal M_C)_{33}}~,\quad b=1-\frac{(\mathcal M_C)_{32}}{(\mathcal M_C)_{33}}~,
\end{gathered}
\end{equation}
the dressing $a,b$ being proportional to the $\overline{\rep{126}}$ admixture. The two
coefficients are nearly equal for a symmetry reason: $\SO{10}$ D-parity (the
$L\leftrightarrow R$ exchange of the Pati--Salam chain) maps $\mathcal O_L\leftrightarrow\mathcal O_R$,
and the dominant mediator---the color triplet of $\rep{10}_H$, a Pati--Salam $(6,1,1)$ and
hence D-parity even---couples to the two bilinears identically, giving $C_R=C_L$ in the
$\rep{10}_H$ limit. Only the $\overline{\rep{126}}_H$ breaks D-parity, so $C_R/C_L$ departs
from unity linearly in the (real) third-generation admixture $|Y_{126}/Y_{10}|\simeq0.06$; we
obtain $|C_R/C_L|\simeq1.07$ with $\phi_{13}\simeq0$.

\paragraph{Dressing into a physical amplitude.}
The bare operators must be dressed into a physical $\Delta B=1$ amplitude by closing the
sfermion line into a one-loop diagram with a gaugino or higgsino~\cite{Ellis:1981tv,Hisano:1992jj}. Although
the gluino carries the largest coupling, the gluino and neutralino loops are flavor-diagonal
and subject to a super-Glashow--Iliopoulos--Maiani (super-GIM) cancellation~\cite{Dimopoulos:1981dw,Hisano:1992jj}: they vanish for exactly degenerate
squarks and switch on only with inter-generation squark splitting. In the universal spectrum
the squarks are degenerate up to small RG running, so these contributions
are strongly suppressed and the \emph{chargino} dressing dominates---a wino loop for
$\mathcal O_L$ and a higgsino loop for $\mathcal O_R$. Both chargino dressings are built from
the Goto--Nihei one-loop function~\cite{Goto:1998qg,Fukuyama:2016vgi}
\begin{equation}
\label{eq:loopF}
F(m_a,m_b)=\frac{\ln(m_a^{2}/m_b^{2})}{m_a^{2}-m_b^{2}}\;\xrightarrow{\,m_a\to m_b\,}\;\frac{1}{m_a^{2}}
~,
\end{equation}
evaluated on the two internal sfermion masses, and read
\begin{align}
\label{eq:dressDX}
D_L
&=
\frac{g_2^{2}}{16\pi^{2}}
\left( \tan\beta+\frac{1}{\tan\beta} \right)\,M_2\,F(m_{\tilde q},m_{\tilde\ell})
~, \\
D_R
&=
\frac{y_t y_\tau}{16\pi^{2}}
\left( \tan\beta+\frac{1}{\tan\beta} \right)^{2}\,\mu\,F(m_{\tilde t},m_{\tilde\tau})
~.
\end{align}
Here $g_2$ is the $SU(2)_L$ coupling, $M_2$ ($\mu$) the wino (higgsino) chargino mass, and
$y_{t,\tau}$ the loop Yukawa vertices. The wino piece carries one power of the chirality
factor $(\tan\beta+1/\tan\beta)$ and the higgsino piece two, so the right-handed dressing is
more $\tan\beta$-enhanced.

\paragraph{Decay width.}
The physical decay amplitudes are $\mathcal A_{LLLL}=\eta_L\,D_L\,\mathcal A_L^{(0)}$ and
$\mathcal A_{RRRR}=\eta_R\,D_R\,\mathcal A_R^{(0)}$, with $\mathcal A_X^{(0)}$ the Cabibbo-weighted
($\lambda_C\simeq0.22$) combination of the Wilson coefficients $C_X$~\cite{Fukuyama:2016vgi}. The decay
width is the sum of the two contributions~\cite{Goto:1998qg,Fukuyama:2016vgi},
\begin{equation}
\label{eq:dim5width}
\begin{gathered}
  \Gamma(p\to K^+\bar\nu)=\Gamma_L+\Gamma_R
  ~, \qquad
  \Gamma_X=\frac{m_p}{32\pi f_\pi^{2}}\,|\beta_H|^{2}\,(A_{\rm LD}A_{\rm SD})^{2}\,|\mathcal A_{XXXX}|^{2}
  ~,
\end{gathered}
\end{equation}
with $m_p$ the proton mass, $f_\pi$ the pion decay constant, $\beta_H$ the lattice hadronic
matrix element of the three-quark proton-to-vacuum operator (dimension GeV$^{3}$), and
$A_{\rm LD},A_{\rm SD}$ the long- and short-distance quantum chromodynamics (QCD) renormalization factors. The color triplet mass band that normalizes the amplitude is set by two-loop gauge coupling running~\cite{Hisano:1992jj}. The long- and short-distance QCD factors are $A_{\rm LD}=0.32$ and $A_{\rm SD}=0.93$~\cite{Fukuyama:2004pb}, common to both chiralities, and $\beta_H=0.0144~\text{GeV}^3$ is taken from Ref.~\cite{Aoki:2017puj}. The factors $\eta_L$ and $\eta_R$ introduced above carry the dimension-five Wilson coefficients from $\MGUT$ down to the SUSY-breaking scale $M_{\rm SUSY}$, and they are evaluated separately for the two operators~\cite{Goto:1998qg},
\begin{equation}
\label{eq:etaLR}
\eta_X=\prod_{i=1}^{3}\left[\frac{\alpha_i(\MGUT)}{\alpha_i(M_{\rm SUSY})}\right]^{-c^X_i/2b_i}\,\exp\!\left[-\frac{1}{16\pi^2}\!\int_{\ln M_{\rm SUSY}}^{\ln \MGUT}\!\!\!\mathcal{I}_X\,dt\right]
~,\end{equation}
where $b_i$ are the one-loop MSSM beta-function coefficients, $c^L=(2/5,\,6,\,8)$ and $c^R=(12/5,\,0,\,8)$ for $U(1)_Y$, $SU(2)_L$ and $SU(3)_c$, and the Yukawa insertions $\mathcal{I}_L=y_\tau^2$ and $\mathcal{I}_R=2y_t^2+2y_\tau^2$. The two differ for a group-theoretic reason: the right-handed operator $u^c u^c d^c e^c$ carries no $SU(2)_L$ charge, so $c^R_2=0$ and it does not have the weak running that acts on $QQQL$ through $c^L_2=6$, and it picks up two additional top-Yukawa insertions. At $M_{\rm SUSY}=1$~TeV, this gives $\eta_L=5.11$ and $\eta_R=2.88$ at $\tb=5$, so the left-handed coefficient is enhanced by a factor $\eta_L/\eta_R\simeq1.8$ relative to the right-handed one.

Because $\kappa$ is fixed by the calibration point, the size of this effect is best quoted with $\kappa$ recalibrated for each choice. Doing so, replacing the two separate factors by a single common one lowers the proton decay bound at $\tb=2$ from $\mzero\gtrsim11.0$~TeV to $9.3$~TeV, an $18\%$ shift and not a refinement that can be dropped.

The quark mixing that enters the Wilson coefficients is likewise taken at the unification scale rather than at the weak scale. Running the CKM matrix up through the MSSM~\cite{Antusch:2013jca} suppresses the two third-generation angles by about $7\%$, $\sin\theta_{13}: 0.00369\to0.00342$ and $\sin\theta_{23}: 0.0405\to0.0375$, while $\theta_{12}$ is stable under the running; the phase convention $\delta_{13}=90^\circ$ is unchanged. Its net effect on the bound is small once $\kappa$ is recalibrated, $11.1\to11.0$~TeV at $\tb=2$, because the same suppression acts on the calibration point and on the physical point.  We use the unification-scale values because that is the scale at which the Wilson coefficients are generated. Both refinements are included in every number quoted below. Through
$D_X$ and $F$, each piece inherits the parametric scaling
\begin{equation}
\label{eq:dim5rate}
\begin{gathered}
  \Gamma_X
  \propto
  \frac{(m_\chi^{X})^{2}}{\mzero^{4}\,\MHC^{2}}\,
  \left(\tan\beta+\frac{1}{\tan\beta}\right)^{2p_X}\,
  \left| \mathcal A_X^{(0)} \right|^{2}
  ~, \qquad
  \mbox{with }~ p_L=1,\;\; p_R=2
  ~,
\end{gathered}
\end{equation}
where $m_\chi^{X}=M_2,\mu$ and the heavy-sfermion limit $F\sim1/\mzero^{2}$. The factor common to both, $1/(\mzero^{4}\MHC^{2})$, makes the lifetime rise as the fourth power of the scalar mass; this is what turns the proton lifetime limit into a \emph{lower} bound on $\mzero$ that strengthens with $\tan\beta$, the strongest lower bound in the model and the reason a light, natural supersymmetric spectrum is excluded. The rate \cref{eq:dim5rate} is evaluated point by point over the $(\tan\beta,\mzero)$ plane and over the electroweakino dressing mass, so the resulting bound on $\mzero$ is a curve rather than a single point. The form $\Gamma_X\propto (m_\chi^{X})^{2}/\mzero^{4}$, and with it $\mzero\propto (m_\chi^{X})^{1/2}$, is the \emph{asymptotic} behavior reached only once the electroweak mixing in the chargino mass matrix is negligible. It is the mass insertion $M_2=\mu$ that the dressing amplitude is proportional to, not the physical chargino mass $m_\chi^{X}$.  For $M_2=\mu$, the two chargino eigenvalues remain split by $\sqrt{2}\,m_W(\sin\beta+\cos\beta)\simeq134$~GeV. At $m_\chi^{X}=0.1$~TeV, the higgsino dressing mass is therefore $2.3$ times $m_\chi^{X}$, which raises the rate above a strict $(m_\chi^{X})^{2}$ law and flattens the bound to $\mzero\propto (m_\chi^{X})^{0.45}$.  The quoted values come from the point-by-point evaluation, not from the asymptotic form. At $\tan\beta=5$ the bound moves from about $5.8$ to $26.8$~TeV as $m_\chi^{X}$ ranges over $0.1$--$3$~TeV (the light orange band in \cref{fig:bnd_tanb}), and we quote the reference value $m_\chi^{X}\simeq1$~TeV throughout.

We collect here, in one place, everything that moves this bound. Every number in this paragraph is quoted at $\tb=2$ and, where a dressing mass is not stated, at $m_\chi=1$~TeV, $m_\chi$ being the mass of the lighter chargino. Two are theoretical uncertainties. The overall normalization of the dimension-five rate, the constant $\kappa$ set by the calibration reference point, carries a factor $\lesssim1.7$, which through $\mzero\propto\tau_p^{1/4}$ is $\pm14\%$ on $\mzero$, a band of about $[9.6,12.5]$~TeV around the $11.0$~TeV central value. The color triplet mass band of \cref{sec:theory_unif} gives a further factor $\simeq 1.6$ through $\mzero\propto\MHC^{1/2}$. The lattice matrix element $\beta_H$ and the residual spread over the color triplet phase $\phi_{13}$, which moves the $\tb=2$ bound within $[9.9,11.6]$~TeV, are both subdominant. The dressing mass $m_\chi$ is by contrast an input rather than an uncertainty, and it is the largest single lever on the number.  At $\tb=2$ the bound runs
\begin{equation}
\label{eq:mchirange}
\mzero \gtrsim 4.1,~~11.0,~~18.4~\text{TeV}
\qquad\text{for}\qquad
m_\chi = 0.1,~~1,~~3~\text{TeV}
~,
\end{equation}
well described by $\mzero\propto m_\chi^{0.45}$ over that range.  The LEP chargino limit already allows the lowest of these, $m_\chi\simeq0.1$~TeV.  The corresponding numbers at $\tb=5$, $5.8$ and $26.8$~TeV at the two ends, are larger throughout because the bound rises with $\tb$.  So every quoted value of $\mzero$ carries the reference choice $m_\chi\simeq1$~TeV with it.

\paragraph{Left/right balance.}
Because $C_R/C_L\simeq1$, the amplitude ratio
\begin{equation}
  \frac{|\mathcal A_{RRRR}|}{|\mathcal A_{LLLL}|}
  =\frac{D_R\,|\mathcal A_R^{(0)}|}{D_L\,|\mathcal A_L^{(0)}|}
\end{equation}
has its dominant growth from the extra power of the chirality factor $(\tan\beta+1/\tan\beta)$ in $D_R/D_L$, with a milder further rise from the Yukawa and Wilson-coefficient weighting (so the computed ratio grows somewhat faster than $(\tan\beta+1/\tan\beta)$ alone). It is independent
of $\mzero$, grows with $\tan\beta$, and crosses unity at $\tan\beta\simeq7.6$
(\cref{tab:LRsplit}). Across the allowed window $\tan\beta \lesssim 8.0$ (to be seen in \cref{sec:walls_mh}), the
left-handed piece dominates below $\tan\beta\simeq7.6$ and the right-handed piece
above.  Because this ratio varies by roughly one order of magnitude over the allowed window, neither chirality amplitude of the $p\to K^+\bar\nu$ decay can be neglected, and the customary wino-dressed-only estimate is inadequate here. Several observations follow. As the two pieces add coherently and, since the real third-generation Yukawas force $\phi_{13}\simeq0$ (below), constructively, their comparable size \emph{raises} the total rate above a left-handed-only calculation and strengthens the proton-decay lower bound on $\mzero$; dropping the right-handed operator would instead give a too-loose bound. The $\mathcal{O}(1)$ ratio is also what makes the relative phase $\phi_{13}$ physically consequential: the interference term is comparable to the diagonal ones, so although the \emph{total} rate shifts by only tens of percent under $\phi_{13}\to\pi$ (diluted because the $\bar\nu_\mu,\bar\nu_e$ channels carry the left-handed operator alone), the $\bar\nu_\tau$ partial rate is strongly phase-sensitive. Because the two operators carry different flavor structures, the decay \emph{pattern} is $\tan\beta$-dependent, and the region above the crossing is right-handed dominated. We do not, however, put this forward as a signature to be tested. The right-handed region inside the allowed window runs only from the crossing at $\tan\beta\simeq7.6$ to the upper limit at $8.0$, whereas that upper limit carries the asymmetric uncertainty $^{+5.5}_{-2.0}$ of \cref{eq:wedge}. The interval is therefore contained well within the uncertainty of its own boundary, and no measurement of the branching composition could resolve it. Finally, the comparable right-handed piece is a fingerprint of the $\overline{\rep{126}}$, whose couplings generate both $C_R$ and the Majorana/seesaw structure, tying the high-$\tan\beta$ proton-decay edge to the same $\tan\beta$-enhanced down-quark and charged-lepton Yukawas that fix $m_b$ and $m_\tau$. The same trend appears in a strongly hierarchical sfermion spectrum, where the light-generation left-handed loops decouple and the right-handed piece dominates throughout.

\begin{table}[htbp]
\centering
\caption[Left- vs right-handed proton amplitude ratio.]{The ratio
$|\mathcal{A}_{RRRR}|/|\mathcal{A}_{LLLL}|$ of the right-handed (higgsino-dressed) to
left-handed (wino-dressed) $p\to K^+\bar\nu$ amplitudes, which is independent of $\mzero$. It
grows with $\tb$ and crosses unity at $\tb\simeq7.6$, below (above) which the left-handed (right-handed) piece dominates.}
\label{tab:LRsplit}
\bigskip
\begin{tabular}{cccc}
\toprule
$\tb$ & $2.5$ & $6$ & $10$ \\
\midrule
$|\mathcal{A}_{RRRR}|/|\mathcal{A}_{LLLL}|$ & ~$0.146$~ & ~$0.672$~ & ~$1.795$~ 
\\
\bottomrule
\end{tabular}
\end{table}

\begin{figure}[htbp]
\centering
\includegraphics[width=0.6\columnwidth]{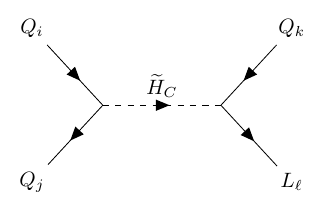}
\caption[Dimension-five colored higgsino exchange for $p\to K^+\bar\nu$.]{The dimension-five
amplitude that drives $p\to K^+\bar\nu$ in minimal supersymmetric $\SO{10}$. The colored
higgsino is exchanged between the external matter superfield legs, generating
the $B$-violating four-fermion operator $\mathcal{O}_L=QQQL$ of \cref{eq:dim5ops}.
(The squark$\to$gaugino/higgsino loop dressing that produces the physical $\Delta B=1$
amplitude is not drawn.)}
\label{fig:dim5}
\end{figure}

\paragraph{Phase, modes, and experimental limits.}
The two pieces add coherently in the $\bar\nu_\tau$ channel (the $\bar\nu_\mu,\bar\nu_e$
channels carry only $\mathcal O_L$) with a relative phase $\phi_{13}$. We derive this
phase rather than scan over it: the measured third-generation $b,\tau$ Yukawas are real,
forcing a real relative sign, so the realized $\phi_{13} \simeq 0$ and the interference is
\emph{constructive}; flipping $\phi_{13} \to \pi$ shifts the total rate by only tens of percent, so the
bound is robust against it. The most important mode is $p\to K^+\bar\nu$ (the strange final state is
favored because the dominant operator carries a second-generation index and the kaon is
kinematically accessible).  As a cross-check, we also compute $p\to\mu^+K^0$ from the same
coefficients and overall normalization~\cite{Goto:1998qg,Fukuyama:2016vgi}. We impose the current limits
\begin{equation}
\label{eq:protoncuts}
\begin{gathered}
  \tau_p(p\to K^+\bar\nu)>6.61\times10^{33}\,\text{yr~\cite{PDG26}}
  ~, \\
  \tau_p(p\to\mu^+K^0)>3.6\times10^{33}\,\text{yr~\cite{Super-Kamiokande:2022egr}}
  ~,
\end{gathered}
\end{equation}
with the first superseding the earlier Super-Kamiokande value $>5.9\times10^{33}\,$yr~\cite{Super-Kamiokande:2014otb}. We write $\tau_p=\kappa/\Gamma_{\rm shape}$, computing
the parameter dependence $\Gamma_{\rm shape}$ and fixing the single overall constant
$\kappa$ once to the published Goto--Nihei rate (a factor-of-$\sim2$
calibration~\cite{Goto:1998qg}). We fix the overall constant $\kappa$ so that our computation gives $\tau_p = 3 \times 10^{31}$~yr at the
Goto--Nihei reference point ($\mzero=1$~TeV, $M_2=125~\text{GeV}$, $\mu=500~\text{GeV}$, $\tb=2.5$). \cref{fig:dim5} shows the mechanism. This absolute normalization is calibrated to the Goto--Nihei rate and uses the single effective $\MHC$; it does not invoke the color triplet doublet-mixing factors of specific complete-model constructions. Because $\kappa$ is fixed in this way, the Goto--Nihei point is a calibration and not a test of our amplitude.  What the calibration does fix externally is the overall scale of the rate.  The dependence on $\mzero$, $\tan\beta$ and the electroweakino masses, and the relative weight of the left- and right-handed operators, are ours. The dressed amplitude has been evaluated using {\tt Python} and {\tt Mathematica} codes, which agree at the three points $(\tb,\mzero/{\rm TeV})=(2,12)$, $(5,20)$ and $(8,25)$ inside the allowed region to better than $10^{-8}$ (see \cref{tab:codebench}).  The two codes are written based on the same formulas, as a check of the coding.  No public code computes this observable for the minimal $\SO{10}$ color triplet sector, and the amplitude that gives the main result therefore has no check from outside the work of the authors.

One further comparison suggests itself, and we explain why we do not make it.  Reproducing the published Goto--Nihei lifetimes at parameter points \emph{away} from the calibration point would be a real test rather than a calibration.  Their calculation, however, is for minimal $\SU{5}$ supergravity.  The hadronic matrix element $\beta_H$, the QCD renormalization factors $A_{\rm LD}A_{\rm SD}$, and the dressing loop function $F(m_a,m_b)$ of \cref{eq:loopF} are common to the two models, but the Wilson coefficients are not.  Ours follow from the $\rep{10}\oplus\overline{\rep{126}}$ fit of \cref{eq:CL,eq:CR} and differ from the $\SU{5}$ ones both in flavor structure and in the chirality ratio, which is $|C_R/C_L|\simeq1.07$ here and is generated with a different weight in minimal $\SU{5}$.  Evaluating our amplitude at their parameter points would therefore mix a difference of implementation with a difference of model, and neither outcome could be read cleanly.  A meaningful comparison would be outside the scope of the present analysis.

The dimension-five rate scales as $\Gamma\propto\MHC^{-2}$, so a longer proton lifetime demands
a heavier colored higgsino: proton decay therefore bounds $\MHC$ from \emph{below}. With grand unification
bounding $\MHC$ from above and proton decay raising it from below, $\MHC$ is pinched from both
sides, leaving $\mzero$ as the only free quantity, as shown by the red region for $\tb \in [2,8]$ in \Cref{fig:bnd_mhc}.  In the ``central'' case of $\tb = 5$ given by the thick red curve, the allowed strip first opens at $\mzero \simeq 15.3$~TeV (marked by the vertical dotted line), where the descending proton decay lower bound meets the upper edge of the grand unification band, and widens to the full band for $\mzero \gtrsim 22.3$~TeV.  One cannot raise $\MHC$ to evade the proton bound, because the same $\MHC$ also sets the unification condition.

\subsection{Dimension-six proton decay}\label{sec:walls_dim6}

\begin{figure}[htbp]
\centering
\includegraphics[width=0.8\columnwidth]{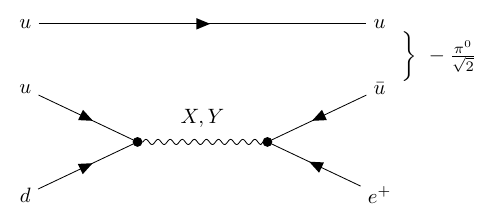}
\caption[The dimension-six gauge proton decay mode $p\to e^+\pi^0$.]{The
dimension-six $p\to e^+\pi^0$ decay, mediated by the heavy $X,Y$
grand unified gauge bosons and suppressed by $1/\MGUT^{2}$, as computed in \cref{eq:dim6rate}. At
$\MGUT=1.43\times10^{16}$~GeV found in \cref{sec:theory_unif}, the lifetime is $3.3\times10^{35}$~yr, far
longer than the dimension-five mode of \cref{fig:dim5}.}
\label{fig:dim6}
\end{figure}

Another grand unified proton decay mode is the \emph{dimension-six} channel
$p\to e^+\pi^0$, mediated by tree-level exchange of the heavy $X,Y$ gauge bosons that
acquire mass $\sim \MGUT$ when $\SO{10}$ breaks, as shown in \cref{fig:dim6}. Integrating them out gives
dimension-six four-fermion operators suppressed by \emph{two} powers of the heavy
scale, so the rate scales as
\begin{equation}
\label{eq:dim6rate}
\begin{gathered}
  \Gamma(p\to e^+\pi^0)
  \sim
  \frac{g_{\rm GUT}^{4}}{\MGUT^{4}}\,m_p^{5}
  ~, \qquad
  \tau_p(p\to e^+\pi^0)
  \sim
  10^{35}\text{--}10^{36}\,\text{yr}
  ~,
\end{gathered}
\end{equation}
where $g_{\rm GUT}$ is the unified gauge coupling and the numerical estimate uses
the unification scale of \cref{sec:theory_unif}, $\MGUT= 1.43\times10^{16}$~GeV, and $\alpha_{\rm GUT}=g_{\rm GUT}^2/4\pi\approx1/24$, together with the lattice hadronic matrix element.  The two chiral structures generated by gauge boson exchange do not carry the same weight: with the operators in the form given in Ref.~\cite{Hisano:1992jj}, the CKM multiplicity factor $(1+|V_{ud}|^{2})$ multiplies the structure with a left-handed diquark alone.  Hence, the squared amplitude carries $|W_0|^{2}\big[A_1^{2}+A_2^{2}(1+|V_{ud}|^{2})^{2}\big]$ rather than a common factor, with $W_0=0.131~\text{GeV}^{2}$ being the lattice matrix element of Ref.~\cite{Aoki:2017puj} (equal for the two structures by parity) and $A_{1,2}$ being the renormalization factors from $\MGUT$ down to $2$~GeV.  Taking $A_1=2.78$ and $A_2=2.91$~\cite{Haba:2019bzs}, we obtain
\begin{equation}
\label{eq:dim6tau}
\tau_p(p\to e^+\pi^0)=5.15\times10^{35}~\text{yr}~~[\SU{5}]
~, \qquad
3.34\times10^{35}~\text{yr}~~[\SO{10}]
~.
\end{equation}
The $\SO{10}$ value is shorter because the additional $(3,2,1/6)$ leptoquark gauge boson of the left--right completion feeds the structure that the $\SU{5}$ bosons leave with unit weight.  In the limit of degenerate heavy gauge bosons, both structures then carry the same factor $(1+|V_{ud}|^{2})^{2}$, so the squared amplitude becomes $|W_0|^{2}(A_1^{2}+A_2^{2})(1+|V_{ud}|^{2})^{2}$ and the two rates stand in the ratio
\begin{equation}
\frac{\Gamma_{\SO{10}}}{\Gamma_{\SU{5}}}
=\frac{(A_1^{2}+A_2^{2})\,(1+|V_{ud}|^{2})^{2}}{A_1^{2}+A_2^{2}(1+|V_{ud}|^{2})^{2}}
=1.54
~,
\end{equation}
which is the ratio of the $\SU{5}$ lifetime to the $\SO{10}$ one in \cref{eq:dim6tau}.  Only when the two renormalization factors are taken equal does this reduce to $2(1+|V_{ud}|^{2})^{2}/[1+(1+|V_{ud}|^{2})^{2}]=1.58$, so the realized enhancement at $A_1=2.78$ and $A_2=2.91$ is a factor of $1.54$.  The structure promoted by the extra boson is the one carrying the smaller factor $A_1$, and promoting it therefore gains less than it would if the two factors coincided.  If the second boson is lighter, the weight of the promoted structure grows from $(1+|V_{ud}|^{2})^{2}$ to $(1+|V_{ud}|^{2}(M_X/M_Q)^{2})^{2}$, so that the ratio above becomes $[A_1^{2}(1+|V_{ud}|^{2}(M_X/M_Q)^{2})^{2}+A_2^{2}(1+|V_{ud}|^{2})^{2}]/[A_1^{2}+A_2^{2}(1+|V_{ud}|^{2})^{2}]$, which rises steeply: it is $5.3$ at $M_X/M_Q=2$ and $18$ at $M_X/M_Q=3$, where the $\SO{10}$ lifetime would fall to $2.8\times10^{34}$~yr, only $1.2$ times the present limit.  Therefore, degeneracy is an assumption rather than a prediction here.  The normalization is checked externally against the detailed minimal $\SU{5}$ calculation of Ref.~\cite{Hisano:1992jj}: evaluated with their inputs, our expression reproduces their published $\tau_p=1.1\times10^{36}\,(M_V/10^{16}~\text{GeV})^{4}(0.003~\text{GeV}^3/\alpha_H)^{2}$~yr to better than $0.1\%$.  The residual uncertainty is dominated by the renormalization factors, which move the lifetime by about $25\%$ across the range quoted in the literature.  This is the channel that originally motivated
proton decay searches, and it would dominate in a non-supersymmetric GUT. Here the
$1/\MGUT^{2}$ suppression makes it orders of magnitude slower than the dimension-five
mode of \cref{eq:dim5rate}, and $p\to e^+\pi^0$ lies well below the present
Super-Kamiokande sensitivity to set a bound: the $\SO{10}$ value of \cref{eq:dim6tau} is $14$ times the current limit $\tau_p>2.4\times10^{34}$~yr~\cite{Super-Kamiokande:2020wjk} and $3$ times the projected Hyper-Kamiokande reach of $\sim10^{35}$~yr, and it stays above the current limit by a factor of $7$ even when every choice that shortens the lifetime is made at once. We compute it for completeness
and compare the two channels in \cref{sec:scan}, but do not impose it as a
constraint.

\subsection{The Higgs mass imposing an upper bound on \texorpdfstring{$\mzero$}{m0}}\label{sec:walls_mh}

In the MSSM, the mass of the lightest $CP$-even Higgs boson is not an input but a
\emph{prediction}. Its quartic self-coupling is not free but fixed by the electroweak
gauge couplings (the $D$-term potential), which at tree level bounds it from above by
\begin{equation}
\label{eq:mhtree}
  m_h^{\rm tree}\le m_Z\,|\cos2\beta|\;\le\;m_Z 
  ~.
\end{equation}
Hence, a tree-level MSSM Higgs mass cannot exceed $m_Z \simeq 91$~GeV. The gap up to the
observed $125$~GeV must be supplied by radiative corrections, dominated by top/stop
loops that grow logarithmically with the stop mass plus a contribution from the stop
mixing $X_t$~\cite{Okada:1990vk,Ellis:1990nz,Haber:1990aw}:
\begin{equation}
\label{eq:mhloop}
  m_h^{2}
  \simeq 
  m_Z^{2}\cos^{2}2\beta
  +\frac{3\,m_t^{4}}{2\pi^{2}v^{2}}\!\left[\ln\frac{m_{\tilde t}^{2}}{m_t^{2}}
  +\frac{X_t^{2}}{m_{\tilde t}^{2}}\!\left(1-\frac{X_t^{2}}{12\,m_{\tilde t}^{2}}\right)\right]
  ~.
\end{equation}
To reach $125$~GeV, one needs either heavy stops or large mixing. \Cref{eq:mhloop} is the one-loop leading-logarithm result, shown here for intuition; for the multi-tens-of-TeV stops in this model, its logarithm grows too large, so the analysis instead uses the resummed Higgs mass calibrated to \texttt{FeynHiggs}, which levels off as the stop mass rises. That leveling-off keeps the low-$\tb$ part of the allowed region open and places the upper limit on $\tb$ at $8.0$.

Because the correction in \cref{eq:mhloop} is bounded, this places an upper bound on
$\mzero$: too heavy a stop, and hence too large an $\mzero$, drives the loop above
$125$~GeV, and the stop mixing $X_t$ cannot compensate, since it only raises $m_h$.
The tree piece $\propto|\cos2\beta|$ saturates by $\tb\sim5$--$10$.  So beyond that, larger $\tb$ raises the tree-level mass no further and the loop contribution must fill the gap; the upper bound on $\mzero$ therefore tightens as $\tb$ grows. It meets the rising proton decay
lower bound of \cref{sec:walls_dim5} at $\tb \simeq 8.0$, which is what makes the
low-to-moderate-$\tb$ a consequence of the data rather than an assumption.  The physically allowed window closes by $\tb\simeq8.0$; the illustrative figures below span $\tb\in[2,25]$, while the global scan samples $\tb\in[2,50]$ (\cref{tab:inputs}).

The precise value of the Higgs mass used in our analysis is the world average $m_h=125.13\pm0.11$~GeV~\cite{PDG26}. We evaluate $m_h$ with an EFT-resummed
treatment: a one-loop leading-log expression calibrated point-by-point to
\texttt{FeynHiggs}~\cite{Heinemeyer:1998yj}, which at the multi-tens-of-TeV stop masses of the
surviving region lies $\sim6$--$9$~GeV below the bare leading-log (the latter omits the
negative two-loop and resummation terms). This calibration sets the upper reach of $\tb$; the bare leading-log, being several GeV too high at these stop masses, would misplace it. This resummed value matches a native \texttt{FeynHiggs} computation and lies about $17$~GeV below the fixed-order \texttt{SPheno~4.0.7}~\cite{Porod:2003um} result at the same heavy-stop spectrum. The fixed-order calculation omits the resummation of the large stop logarithm $\ln(M_S^2/m_t^2)\approx11$ (at $M_S\sim45$~TeV), which the EFT treatment sums, so in \cref{tab:codebench} the \texttt{SPheno} entry is this fixed-order value while the \texttt{FeynHiggs} entry is the EFT-resummed value used in our analysis.  The $\sim17$~GeV separation is therefore the expected fixed-order overshoot at these masses, and we do not treat it as an uncertainty of our Higgs-mass prediction; the uncertainty we assign is the residual $\sim1.5$~GeV theory error of the resummed Higgs mass. The calibration is done point by point against \texttt{FeynHiggs~2.19} over the domain $\tb\in[2,20]$ and $M_S\in[3,36]$~TeV, which contains the entire surviving region.  The $\pm1.5$~GeV uncertainty is built and rounded from:
\begin{equation}
\label{eq:mherr}
\begin{split}
\delta m_h = \big[\,&(0.65)^2_{\rm FH} + (0.72)^2_{\rm calib} + (0.27)^2_{m_t} \\
&+ (0.20)^2_{\alpha_s} + (0.11)^2_{m_h} + (1.0)^2_{\rm match}\,\big]^{1/2} = 1.44~\text{GeV}
~,
\end{split}
\end{equation}
The first term is the uncertainty \texttt{FeynHiggs} itself reports at the $\mzero=45$~TeV spectrum of \cref{tab:codebench} ($0.57$--$0.96$~GeV over the four heavy-stop benchmarks). The second is the largest departure of our calibrated expression from \texttt{FeynHiggs} anywhere in the calibration domain. The next three are parametric, from $m_t$, from $\alpha_s$, and from the measured $m_h$. The last allows for the higher orders missing from the matching of the effective theory at the stop scale, following Ref.~\cite{Bahl:2019hmm}, which finds $\sim0.5$~GeV at small stop mixing.  We take $1.0$~GeV to be conservative.  Propagated through the meeting of the proton and Higgs bounds, this $\sim1.5$~GeV error gives the best-estimate upper limit $\tb\lesssim8.0$ a large, asymmetric systematic uncertainty, quantified in \cref{sec:scan_strict}.  It does not enter the bounding curves of \cref{fig:bnd_tanb}, which are drawn at $m_h=125$~GeV exactly, and only the $\pm1.5$~GeV theory error propagates into the quoted limits.

\subsection{The LEP chargino bound}\label{sec:walls_lep}

In the mini-split spectrum, the electroweakinos are the only light superpartners, and they are the states that dress the proton decay loop of \cref{sec:walls_dim5}. The non-observation of chargino pair production at LEP2 excludes a light chargino by requiring
\begin{equation}
\label{eq:lep}
  m_{\chi^\pm_1}>103.5\,\text{GeV}
\end{equation}
in the general MSSM~\cite{PDG26}. The limit is conditional: it assumes a non-compressed spectrum with standard decays, and softens toward the LEP kinematic reach $\simeq 92$~GeV when the lightest chargino and neutralino are nearly degenerate (compressed higgsino- or wino-like states) or when the sneutrino-mediated $t$-channel interference depletes the production cross section. In the constrained-$\mu$ spectrum, the lighter chargino is wino-like (mass $\sim M_2$) and sits well above the bino LSP, since gaugino-mass unification gives $M_1<M_2$. It therefore decays promptly rather than forming a compressed, long-lived state. In any case, the electroweakinos in the allowed region lie at the TeV scale, far above the LEP reach, so the chargino bound is not binding. The bound removes the corner where the dressing electroweakino would be too light to be consistent with
collider data. It does not constrain $\mzero$, but sets the lower end of the
electroweakino axis of the projection in \cref{sec:walls_panels}.

\subsection{The individual constraints in the parameter space}\label{sec:walls_panels}

\begin{figure}[htbp]
\centering
\includegraphics[width=0.8\columnwidth]{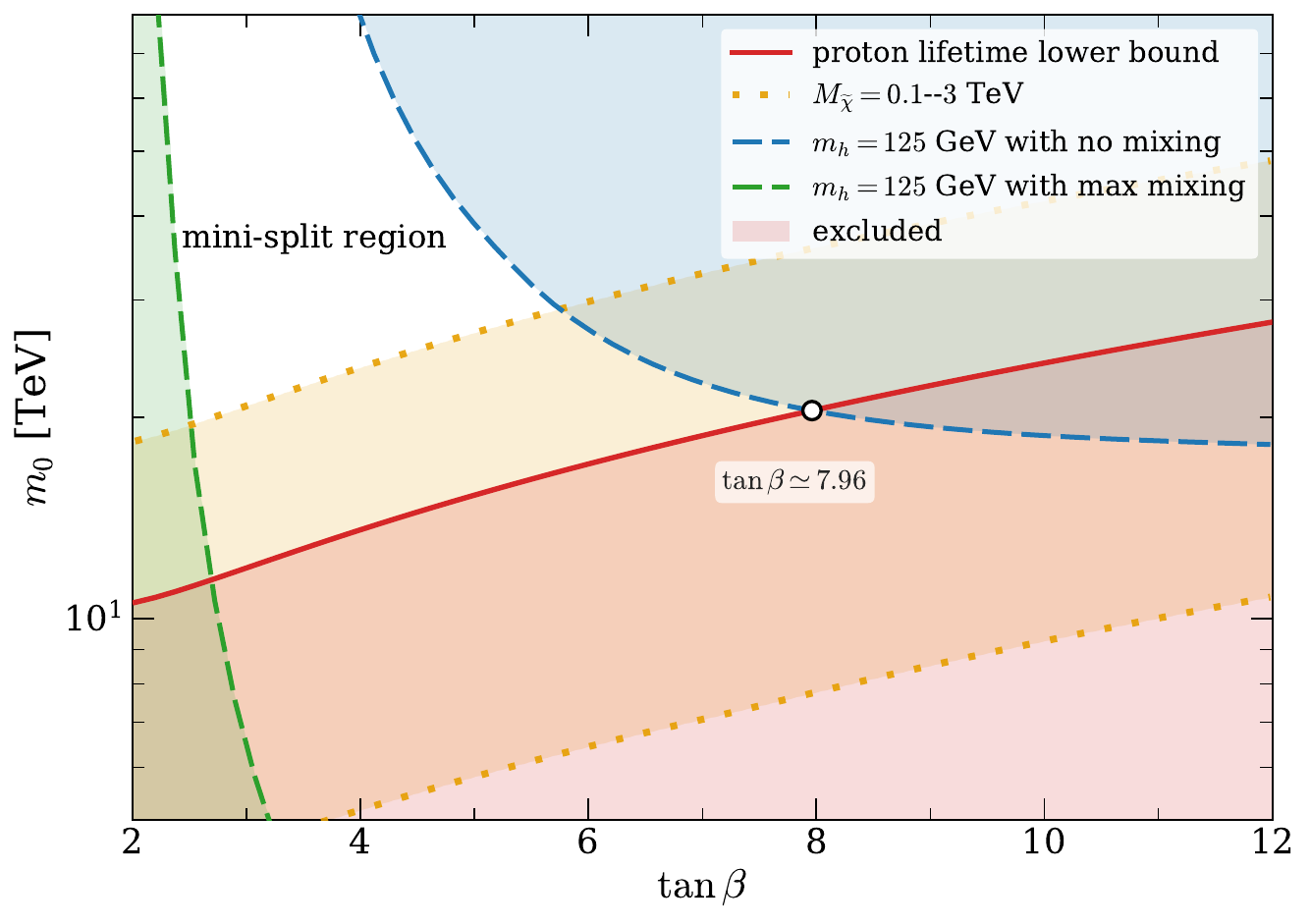}
\caption[The allowed region in the $(\tb,\mzero)$ plane.]{The allowed region in the
$(\tb,\mzero)$ plane. The proton decay lower bound (red solid), shown for the reference dressing mass $\mchi \simeq 1$~TeV, and the $m_h=125$~GeV upper bound (blue dashed) bound the mini-split region, which closes at $\tb \simeq 8.0$ ($\mzero \simeq 20.5$~TeV). The two dashed $m_h$ curves are the edges of the $m_h=125$~GeV band traced as the stop mixing $\Xt$ is varied: the upper edge (blue dashed) is the
largest $\mzero$ reaching $m_h=125$~GeV at zero or minimal stop mixing, and the lower edge (green dashed) is the
smallest $\mzero$ reaching it at maximal mixing $|\Xt|=\sqrt6\,\mzero$.~The region to the left of this green dashed curve (light green) is excluded because there the tree-level contribution $m_h\propto|\cos2\beta|$ is too small for even maximal mixing to reach $125$~GeV, so the Higgs is too light. The shaded light orange region bounded by the two dotted curves is obtained by varying $\mchi=0.1$--$3$~TeV. The shaded pink, light blue, and light green regions are excluded. Here we fix $M_2=\mu=1$~TeV.}
\label{fig:bnd_tanb}
\end{figure}

\begin{figure}[htbp]
\centering
\includegraphics[width=0.8\columnwidth]{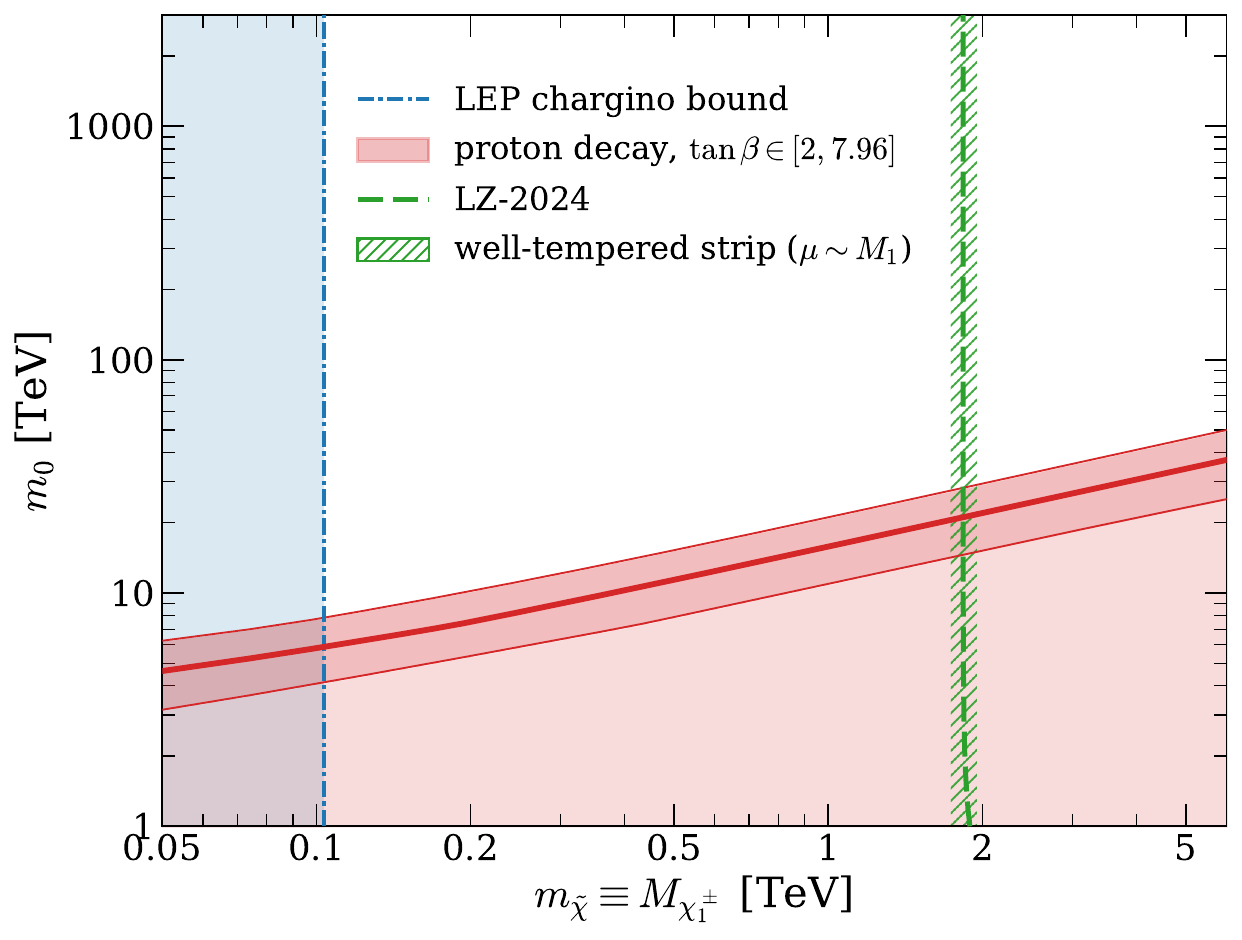}
\caption[Electroweakino and DM bounds in $(\mchi,\mzero)$.]{The
electroweakino projection in the $(\mchi,\mzero)$ plane: the LEP chargino bound
($m_{\chi^\pm_1}>103.5$~GeV) and the proton decay bound, plus the DM direct detection bound. The direct detection exclusion (green hatched region) applies only when the lightest neutralino is the thermal DM (the well-tempered strip; see \cref{sec:dm}). In this plot, we take $\tan\beta=5$ and $M_2=\mu=1$~TeV.  The proton bound uses $\tau_p(p\to K^+\bar\nu)>6.61\times10^{33}$~yr.}
\label{fig:bnd_mchi}
\end{figure}

We now show how these limits bound the $(\tb,\mzero)$ plane, deferring the full global
scan to \cref{sec:scan}. \cref{fig:bnd_tanb} shows the two that determine the allowed
region: the rising proton decay lower bound and the falling $m_h=125$~GeV upper bound on $\mzero$,
which close at $\tb \simeq 8.0$ and $\mzero \simeq 20.5$~TeV, with the lower bound reaching
down to $\mzero \gtrsim 11.0$~TeV at low $\tb$. Gauge unification enters indirectly,
normalizing the proton decay bound through the $\MHC$ band of \cref{sec:theory}, while
flavor and DM (\cref{sec:flavor,sec:dm}) do not constrain this plane. The two edges depend on $\{\mzero,\MHC,\tb\}$ and, through the dressing loop, on the electroweakino mass $\mchi$.  The fixed $M_2=\mu=1$~TeV in \cref{fig:bnd_tanb,fig:bnd_mchi} fixes only the parameter values plotted in each figure, not the edges themselves.  The sole residual dependence of the edges on $m_\chi$ is shown as the $m_\chi = 0.1$--$3$~TeV band.

\cref{fig:bnd_mchi} projects onto the electroweakino plane $(\mchi,\mzero)$, showing
the LEP bound of \cref{eq:lep} and the proton decay bound together with the
direct detection bound to be discussed in \cref{sec:dm}. The proton decay bound is unconditional.  The LEP chargino bound is conditional on the electroweakino spectrum (\cref{sec:walls_lep}), though non-binding here.  The direct detection bound is conditional, applying only to the well-tempered strip and only under the thermal-neutralino DM hypothesis.

\section{Flavor physics and precision observables}\label{sec:flavor}

The flavor- and $CP$-violating observables, together with the muon anomalous magnetic moment, probe structure beyond the universality assumed for the soft scalar masses, and exact universality would switch them all off. Two small but irreducible pieces survive: the RG-induced slepton mixing that drives $\mu\to e\gamma$, and the residual SM $CP$ violation that feeds the EDMs. We compute all of them over the allowed region, and none are restricted in the heavy, $CP$-conserving (real-soft) regime that the proton bound forces, under our universality assumption of the soft scalar masses. The multi-TeV scalars
required by proton decay (\cref{sec:walls}) suppress every supersymmetric loop, and the
real soft terms leave the CKM phase as the sole source of $CP$ violation, keeping the model CKM-safe.

\begin{figure}[htbp]
\centering
\includegraphics[width=0.8\columnwidth]{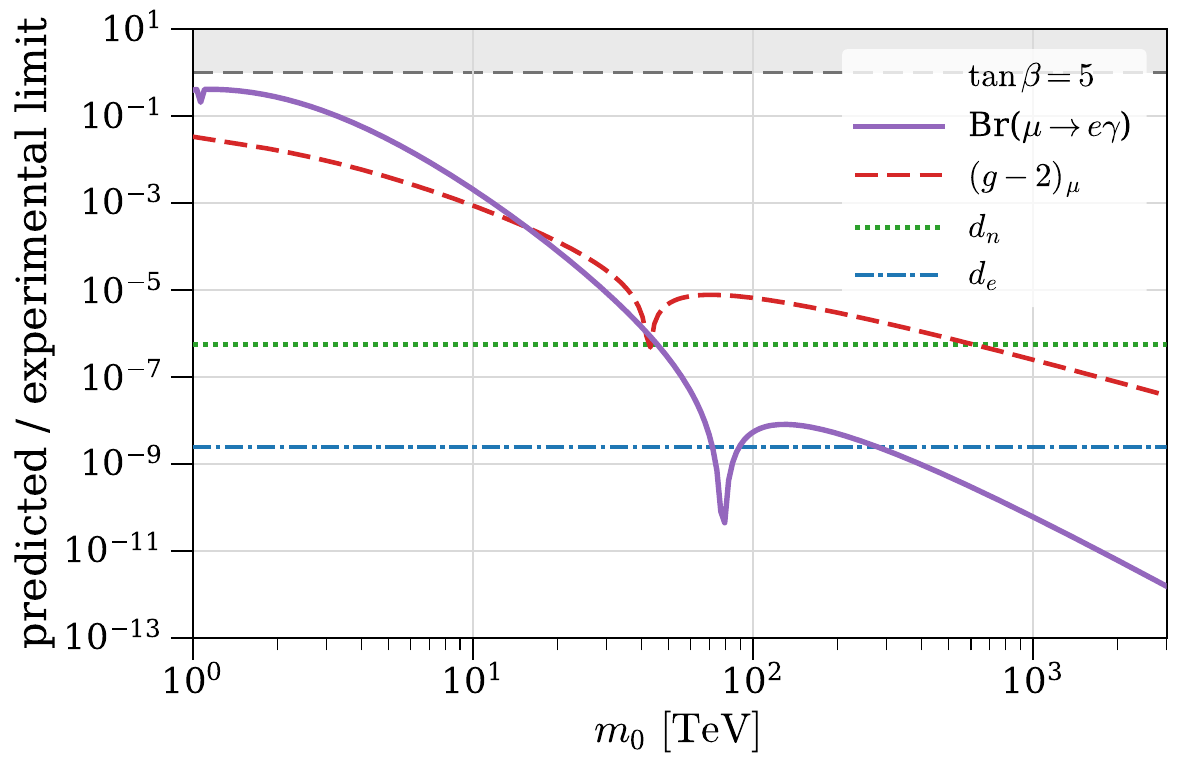}
\caption[Flavor and precision observables: all non-binding.]{Flavor and precision observables ($\mu\to e\gamma$, muon $g-2$, electron and neutron EDM): each sits far below its experimental limit (ratio $\ll1$) across the allowed region. The very heavy scalars suppress the supersymmetric loops, and the real soft sector keeps the model CKM-safe.  We fix $\tb = 5$ in this plot.}
\label{fig:con_nonbinding}
\end{figure}

\paragraph{$\boldsymbol{\mu\to e\gamma}$.}
The lepton flavor-violating (LFV) process, $\mu\to e\gamma$, is not absent even with a 
flavor-blind universal scalar mass $\mzero$. The
neutrino Dirac Yukawa coupling $Y_D$, tied by the unified multiplets to the up-type
Yukawa and hence large, drives an off-diagonal entry in the left-handed slepton
soft-mass matrix through RG running between $\MGUT$ and the
seesaw scale. The process therefore tests the same Yukawa structure that fixes the
fermion masses and the proton decay operators, and the off-diagonal slepton entry is
radiatively irreducible and cannot be removed by a choice of boundary conditions. A
neutralino--slepton (or chargino--sneutrino) loop carrying one $\delta_{12}$ insertion
emits the photon, giving a magnetic-dipole (photon-penguin) transition. The
branching ratio is~\cite{Hisano:1995cp}
\begin{equation}
\label{eq:mueg}
\begin{gathered}
\mathrm{Br}(\mu\to e\gamma)
\simeq
\frac{\alpha^3}{G_F^2\,\mzero^4}\,
\bigl|\delta_{12}\bigr|^2\,\tan^2\beta
~, \qquad
\delta_{12}
\sim
\frac{1}{8\pi^2}\,\bigl(Y_D^\dagger Y_D\bigr)_{12}\,
\ln\!\frac{\MGUT}{M_R}
~,
\end{gathered}
\end{equation}
where $\delta_{12}$ is the radiatively generated left-handed slepton mass insertion, set
by the neutrino Dirac Yukawa $Y_D$ running between $\MGUT$ and the seesaw scale $M_R$.
Because $\mathrm{Br}\propto1/\mzero^{4}$, the heavy scalars demanded by the proton bound
suppress it far below the current MEG~II limit~\cite{MEGII:2025gzr} across the entire allowed
region.  The predicted-to-limit ratio is $\ll1$ everywhere in \cref{fig:con_nonbinding}.

\paragraph{Muon $\boldsymbol{g-2}$.} The supersymmetric contribution to the muon
anomalous magnetic moment $a_\mu^{\rm SUSY}$ (with $a_\mu=(g-2)_\mu/2$) arises at one loop from the
chargino--sneutrino and neutralino--smuon diagrams~\cite{Grifols:1982vx}. Because the smuon and
sneutrino in the loop sit at the heavy scale $\mzero$, the contribution is
\begin{equation}
\label{eq:amu}
a_\mu^{\rm SUSY}
\simeq
\frac{\alpha\,m_\mu^2\,\tan\beta}{8\pi\sin^2\theta_W\,\mzero^2}\,
\,\mathrm{sgn}(\mu) 
~,
\end{equation}
\Cref{eq:amu} is the light-sfermion limit; in the heavy scalar regime here, the contribution decouples one power faster, $\propto M_2\,\mu\,m_\mu^{2}\,\tb/\mzero^{4}$, and the bound quoted next is the full one-loop result evaluated numerically. It is small: over the allowed
region $|a_\mu^{\rm SUSY}|\le7.9\times10^{-13}$, about $1.1\times10^{-3}$ of the present
experimental sensitivity (Brookhaven $E821$~\cite{Muong-2:2006rrc}, now superseded in precision by
Fermilab~\cite{Muong-2:2021ojo}). So the model is SM-like in $(g-2)_\mu$. Under the 2025 lattice-based Muon $g-2$
Theory Initiative evaluation~\cite{Aliberti:2025beg}, which uses lattice QCD for the hadronic
vacuum polarization, the SM prediction sits within $\sim 0.4\,\sigma$ of
experiment (the data-driven $e^+e^-$ evaluation remains in tension), so there is little
or no anomaly to accommodate, and the absence of a sizable supersymmetric shift in the
heavy scalar regime is consistent with this.

\paragraph{Electric dipole moments (electron, neutron).} Supersymmetry with
$CP$-violating soft terms generically induces one-loop EDMs far above the experiment, the
``supersymmetric $CP$ problem'' first identified in Refs.~\cite{Ellis:1982tk,Polchinski:1983zd,Buchmuller:1982ye}. 
Under the universal,
$CP$-conserving soft sector assumed throughout (real $\mzero,\mhalf,A_0$), the only
source of $CP$ violation in the model is the CKM phase. The CKM-induced SM values are
$\sim10^{-38}$ and $\sim10^{-32}\,e$-cm for the electron and neutron EDMs~\cite{Pospelov:2005pr,Czarnecki:1997bu},
respectively, some $\sim 2 \times 10^{-9}$ and $\sim 6 \times 10^{-7}$ of the experimental limits
(electron $|d_e|<4.1\times10^{-30}\,e$-cm~\cite{Roussy:2022cmp}; neutron
$|d_n|<1.8\times10^{-26}\,e$-cm~\cite{Abel:2020pzs}). This follows from
the assumed real soft terms and is not automatic: if an $\mathcal{O}(1)$ $CP$ phase were to
reside instead in the light electroweakino sector, the one-loop sfermion EDM would still
decouple as $1/\mzero^{2}$, but the two-loop Barr--Zee diagram, in which a heavy inner
loop attaches to the fermion through a photon and a Higgs or gauge boson, would not, so the dipole survives the sfermion decoupling. A maximal-phase Barr--Zee estimate,
computed following the split-supersymmetry analysis of Ref.~\cite{Giudice:2005rz} (a maximal $CP$ phase $\sin\phi=1$ on the electroweakino sector, the heavy Higgs states decoupled at $M_A\sim\mzero$), can reach the electron EDM limit, $d_e\sim10^{-27}\,e$-cm for TeV-scale electroweakinos ($M_2,|\mu|\sim0.6$--$0.8$~TeV), above the current bound of $4.1\times10^{-30}\,e$-cm, and so constrains
only a light-gaugino corner. We caution that this Barr--Zee number is an
$\mathcal{O}(1)$-factor estimate, not a full two-loop computation; the conclusion that
EDMs are CKM-safe and non-binding under the real soft sector is unaffected.

\cref{fig:con_nonbinding} collects these flavor and precision observables, each shown as
a ratio to its experimental limit across the allowed region. Every ratio sits far below
unity. The heavy scalars required by the proton bound suppress the loops, and the real
($CP$-conserving) soft terms keep the model CKM-safe. The sharp downward spikes are accidental cancellations: as $\mzero$ increases, the competing chargino- and neutralino-loop contributions to each amplitude shift in relative size and momentarily cross zero. The muon $g-2$ ratio dips where $\Delta a_\mu$ changes sign (around $\mzero\sim45$~TeV; the curve plots $|\Delta a_\mu|$), and the $\mu\to e\gamma$ ratio plunges much more sharply where the flavor-violating dipole amplitude vanishes (around $\mzero\sim80$~TeV), since the branching ratio scales as that amplitude squared.

Being non-binding in this regime, the flavor and precision observables leave the main
result unchanged. The allowed region is set by the theoretical and internal consistency
requirements (\cref{sec:theory}) together with the direct experimental limits: proton
decay, the Higgs mass, and the LEP chargino bound (\cref{sec:walls}).  The DM data
(\cref{sec:dm}) act only in the electroweakino sector.

\section{The electroweakino sector}\label{sec:scan}

The survival wedge from proton decay, the Higgs mass, and unification (\cref{sec:theory,sec:walls}) fix $\{\mzero,\tb,\MHC\}$ but leave the electroweakino sector unspecified. We fix it in two ways that share the same physics modules. Under the universal hypothesis, this sector is \emph{not} free: as anticipated in \cref{sec:theory_mu}, EWSB fixes $|\mu|$ from the high-scale set $\{\mzero,\mhalf,A_0,\tb\}$. We therefore give the constrained-$\mu$ result first (\cref{sec:scan_strict}), as the prediction of the universal model, and then a free-$\mu$ low-scale scan (\cref{sec:scan_headline}) that samples $\mu$ directly, as the minimal NUHM extension of the universal model. The two agree on the survival wedge and differ only in what they imply for the higgsino.

\subsection{The constrained-$\mu$ sector}\label{sec:scan_strict}

Under the universal boundary conditions, $|\mu|$ is not a free input: it is fixed, up to its sign, by the two electroweak-minimization conditions once $\{\mzero,\mhalf,A_0,\tb\}$ are given, so the electroweakino sector is a prediction of the model. We fix the universal set of variables at the GUT scale, integrate the RG equations down, and read $|\mu|$ off the minimization condition.

\begin{figure}[htbp]
\centering
\includegraphics[width=0.8\columnwidth]{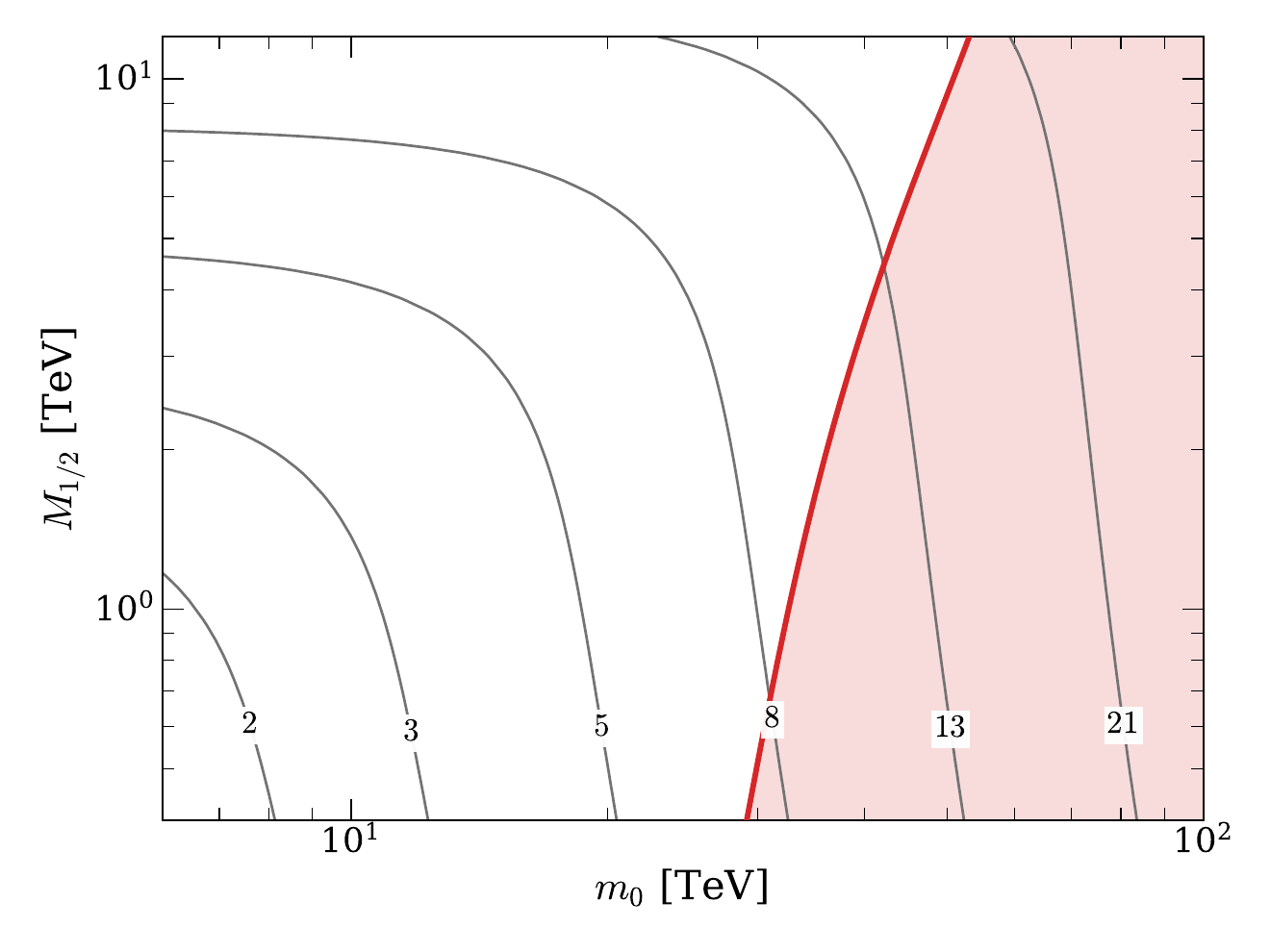}
\caption[The EWSB-determined higgsino mass in the constrained-$\mu$ scenario.]{Contours of
$|\mu|$ (in TeV) as the EWSB output over the $(\mzero,\mhalf)$ input plane at $\tb=5$ and $A_0=0$
from a scan using \texttt{SPheno}:
$|\mu|\geq 1.5$~TeV everywhere, excluding the light-higgsino scenario. The light-red region
passes the proton, Higgs mass, unification, and LEP cuts, giving $|\mu|=7$--$33$~TeV. The Higgs mass uses the EFT resummation calibrated to \texttt{FeynHiggs}~\cite{Heinemeyer:1998yj}, and the proton rate uses the
$\tau_p > 6.61 \times 10^{33}$~yr limit~\cite{PDG26}.}
\label{fig:mu_ewsb}
\end{figure}

\begin{figure}[htbp]
\centering
\includegraphics[width=0.8\columnwidth]{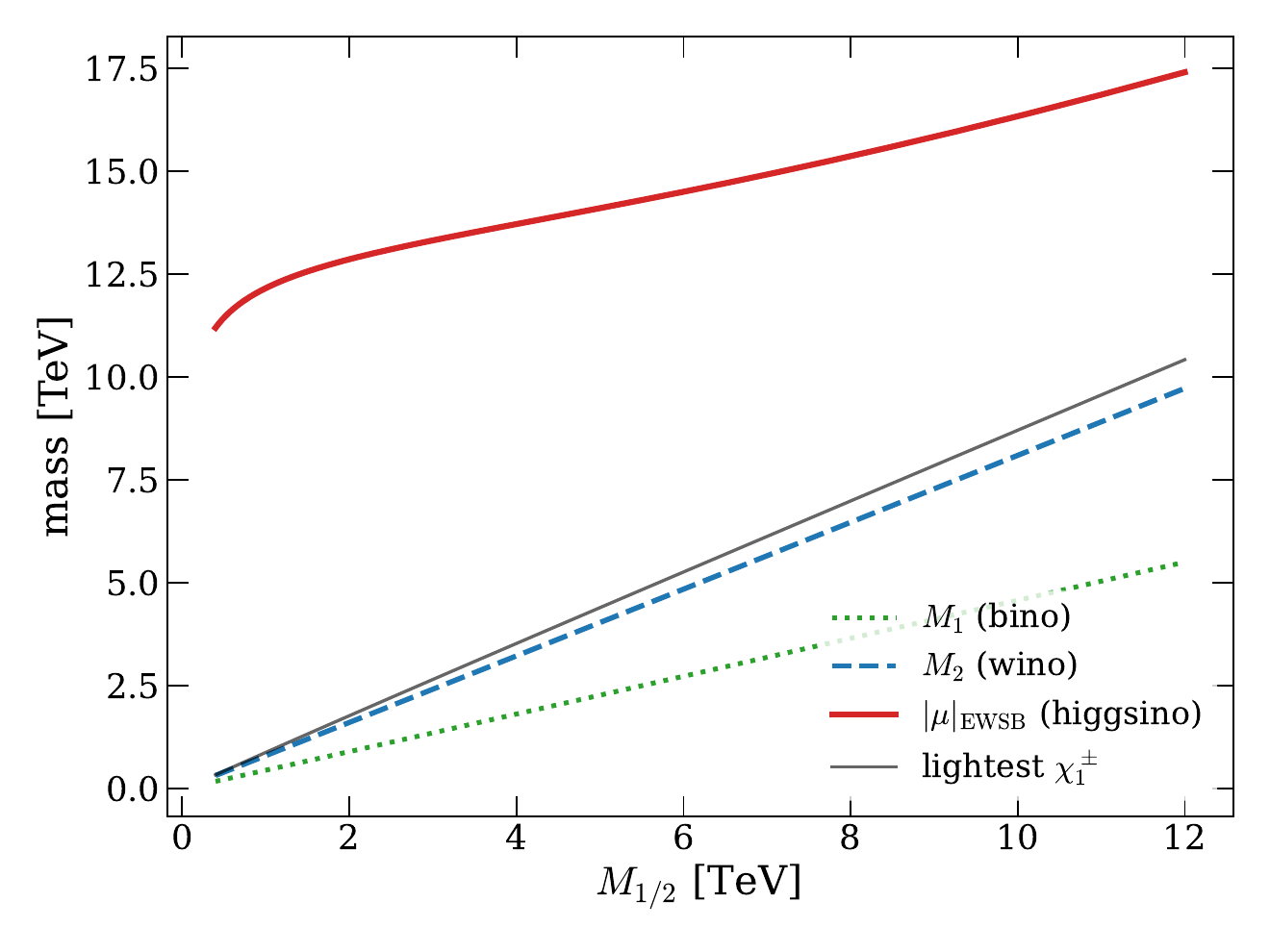}
\caption[The constrained-$\mu$ electroweakino spectrum.]{The constrained-$\mu$ electroweakino spectrum along a surviving $\mzero=45$~TeV slice ($\tb=5$ and $A_0=0$): the EWSB-determined $|\mu|$ lies far above $M_1$ and $M_2$, implying that the higgsino is heavy and the lightest neutralino is the bino, and that the lighter chargino is wino-like.}
\label{fig:spectrum_ewsb}
\end{figure}

\Cref{fig:mu_ewsb} shows the result, computed with \texttt{SPheno} using $\tb=5$ as an example. The EWSB-determined $|\mu|$ is heavy everywhere on the plane (never below $1.5$~TeV) and rises to $7$--$33$~TeV across the light-red surviving region, in which $|\mu|$ always exceeds $M_2$ by a factor of $\mathcal{O}(10)$.  So the higgsino is the heaviest of the three. Since gaugino mass unification gives $M_1<M_2$, the lightest electroweakino is then the bino, not the wino. The spectrum shown in \cref{fig:spectrum_ewsb} along a representative surviving slice makes the ordering explicit: $|\mu|$ sits far above $M_1$ and $M_2$ throughout.  We note that this is a \texttt{SPheno}-based constrained-$\mu$ spectrum result: the other spectrum generators (\texttt{SOFTSUSY~4.1.22}, \texttt{FlexibleSUSY~2.9.0}~\cite{Allanach:2001kg,Athron:2014yba}) reproduce the low-$\tb$ edge but frequently fail to converge in the heavy-scalar region under identical constrained-MSSM inputs as a known problem associated with focus-point supersymmetry~\footnote{This is the expected behavior of a focus-point boundary, where $|\mu|^2$ comes out as a small residual of a large cancellation in the running of the up-type Higgs soft mass. So percent-level differences between codes (in the top-quark Yukawa coupling, the superpartner thresholds, or the matching scale) can readily flip its sign. Where the other generators differ, they return \emph{no} valid EWSB at those points rather than a light higgsino.  Hence, the conclusion that the higgsino is heavy is firm in direction.}.  So we quote the constrained-$\mu$ spectrum as obtained with \texttt{SPheno} rather than as generator-independent. Even where the generators do converge, on the low-$\tb$ edge, the EWSB-determined $|\mu|$ is strongly calculator-dependent: the values from \texttt{SPheno} and \texttt{SOFTSUSY} differ by a median factor of about $4$ (range $2$--$12$), the expected sensitivity of a focus-point $|\mu|^2$ that is a small residual of a large cancellation. We therefore consider the \emph{ordering} $M_1<M_2\ll|\mu|$, a heavy higgsino with a bino lightest neutralino, as the robust output, not the precise value of $|\mu|$. In every converged case, $|\mu|$ comes out multi-TeV and well above $M_2$, and none of our conclusions depends on its exact size.  The $(\tb,\mzero)$ bounds are $\mu$-independent at leading order, and the bino remains a pure, over-abundant thermal relic for any $|\mu|\gg M_1$.

The surviving region is the same narrow mini-split strip whether or not $\mu$ is constrained:
\begin{equation}
\label{eq:wedge}
\begin{split}
&
\tb \lesssim 8.0^{+5.5}_{-2.0}
~, \qquad
\mzero \gtrsim 11.0~\text{TeV}~~(m_\chi = 1~\text{TeV})
~, \\
&
\MHC\in[0.58,1.43] \times 10^{16}~\text{GeV}~~(\text{input})
~,
\end{split}
\end{equation}
with heavy, tens-of-TeV scalars sitting above light gauginos, signaling the high-scale (mini-split)
supersymmetry pattern. The first two entries are results of the analysis, whereas the third is not, since the $\MHC$ band is fixed beforehand by gauge coupling unification (\cref{eq:mhc_band}) and enters as an input that normalizes the proton decay rate. The boundaries are exactly those derived in \cref{sec:walls,sec:theory}
and drawn in the $(\tb,\mzero)$ plane of \cref{fig:bnd_tanb}: the proton (lower) and Higgs-mass (upper) edges meet at $\tb\approx8.0$, with unification fixing $\MHC$ (\cref{fig:bnd_mhc}). At the lowest $\tb$, by contrast, the Higgs mass does not bound $\mzero$ from above because $m_h$ grows only logarithmically with the scalar scale.  At $\tb\simeq2$, the zero-mixing value lies below the $125$~GeV window even for very heavy scalars. Thus, $m_h=125$~GeV is reached only with stop mixing and stays attainable up to arbitrarily large $\mzero$. In other words, the scalars are bounded only by the requirement that they remain below the unification scale at low $\tb$. Of the four families of constraints, only these two draw an edge in this plane; DM and the flavor observables act in the electroweakino and scalar sectors treated in \cref{sec:flavor,sec:dm} and leave the $(\tb,\mzero)$ window untouched.

This constrained-$\mu$ treatment sharpens the bounds rather than weakening them. The three edges of \cref{eq:wedge}, from the proton decay, the $\MHC$ band, and the Higgs mass, depend at leading order only on $\{\mzero,\MHC,\tb\}$ and not on $\mu$.  So the lower bound $\mzero \gtrsim 11.0$~TeV and the upper bound of $\tb \approx 8.0$ stand in either scheme. What the constrained-$\mu$ treatment removes is the light electroweakino, and with it the light-higgsino DM strip (to be seen in \cref{sec:dm}): both are features of treating $\mu$ as a free parameter or of the NUHM extension of \cref{sec:scan_headline}. Strictly speaking, $\mu$ re-enters the three edges at subleading order: through the higgsino dressing of the dimension-five operators, through $\Xt=A_t-\mu\cot\beta$, and through the supersymmetric thresholds in unification.

With the higgsino driven heavy, the lightest superpartner is the bino ($M_1<M_2$). Its DM fate is analyzed in the constrained-$\mu$ scheme of \cref{sec:dm_cmu}. The wino-like states sit above the bino as a near-degenerate pair (split by $\sim160$~MeV~\cite{Ibe:2012sx}) but decay promptly to it, so standard electroweakino searches apply rather than disappearing-track searches. Under the constrained-$\mu$ treatment, the neutralino is thus not a unique thermal candidate, and DM enters as a model-dependent input rather than as part of the core exclusion.

The uncertainties on the $\mzero$ lower bound, and the separate dependence on the dressing mass $m_\chi$, are already collected in \cref{sec:walls_dim5}. The upper bound of $\tb \approx 8.0$ inherits the residual uncertainty of the Higgs mass bound, the least certain of the three: $m_h$ is an EFT--resummed leading logarithm calibrated
to \texttt{FeynHiggs} (the bare leading logarithm, $\sim6$--$9$~GeV too high, would have
closed the window prematurely at $\tb\approx5$), and the residual \texttt{FeynHiggs} theory
error is $\sim1.5$~GeV. This last error dominates the systematic on the upper limit of $\tb$. Because the zero-mixing Higgs mass grows only logarithmically with the stop scale, $m_h^2 \simeq M_Z^2\cos^2 2\beta + (3 m_t^4/2\pi^2 v^2)\,\ln(M_S^2/m_t^2)$, a few GeV of Higgs mass uncertainty translates into a wide range of $M_S$, and hence of the point where the rising proton lower bound meets the falling Higgs upper bound.  This leading-log expression is used only to show the logarithmic sensitivity; the numerical limits quoted are obtained from the EFT-resummed calculation.  Shifting the accepted $m_h$ by $\pm1.5$~GeV moves that meeting point from the best-estimate $\tb\approx8.0$ ($\mzero\approx20.5$~TeV) to $\tb\approx6.0$ ($\mzero\approx17.0$~TeV, at $-1.5$~GeV) and $\tb\approx13.5$ ($\mzero\approx30.6$~TeV, at $+1.5$~GeV).  The shift is strongly one-sided, climbing about $3.7$ units of $\tb$ per GeV on the upper side.  We therefore report $\tb\approx8.0$ as a best estimate carrying this large one-sided theory systematic uncertainty. The proton decay lower bound $\mzero\gtrsim11.0$~TeV and the mini-split character are set by proton decay alone and do not depend on the Higgs window.

\subsection{The free-$\mu$ extension}\label{sec:scan_headline}

A light higgsino is realized only if we relax the universal hypothesis in its Higgs sector: letting the two Higgs soft masses $m^2_{H_u},m^2_{H_d}$ differ from the common sfermion mass $\mzero^2$ at the GUT scale, leading to the NUHM generalization~\cite{Matalliotakis:1994ft,Olechowski:1994gm,Ellis:2002wv}, trades one or two of the electroweak minimization conditions so that $|\mu|$ (and $m_A$) become free low-scale inputs while the sfermions stay universal. This is natural in $\SO{10}$: matter sits in the $\rep{16}$, but the light Higgs doublets descend from the $\rep{10}$ (mixing with the $\overline{\rep{126}}$ and higher representations), and no symmetry forces their soft masses to equal $\mzero$. Keeping the sfermions universal preserves the flavor safety of \cref{sec:flavor}; and the proton, Higgs, and unification edges are unchanged. We therefore read the free-$\mu$ scan below as this minimal NUHM extension of the universal model.

We locate the region with a unified Monte Carlo scan of $10^{7}$ points over the soft-breaking
box of \cref{sec:model} ($\mzero$, $M_2$, $|\mu|$, $\tb$, $\Xt$, $\MHC$), recording a
deterministic pass/fail flag for each of the nine cuts of
\cref{sec:theory,sec:walls,sec:flavor,sec:dm}. The scan reproduces the same survival wedge~\cref{eq:wedge}, an independent confirmation of its $\mu$-independence, and in addition opens the light-higgsino window that the constrained-$\mu$ model closes.  Our amplitude and spectrum are benchmarked against public codes observable by observable, at the points and to the agreement listed in \cref{tab:codebench}, rather than as a single global comparison: the fixed-order and EFT-resummed Higgs mass against \texttt{SPheno} and \texttt{FeynHiggs}, the higgsino relic density and spin-independent cross section against \texttt{micrOMEGAs~6.0}, and the constrained-$\mu$ EWSB spectrum against \texttt{SPheno}, with \texttt{SOFTSUSY}, and \texttt{FlexibleSUSY} reproducing the low-$\tb$ edge~\cite{Allanach:2001kg,Porod:2003um,Athron:2014yba,Heinemeyer:1998yj,Belanger:2001fz}.

\begin{table*}[htbp]
\centering
\caption{Public-code benchmark comparisons underlying the analysis: for each observable, the benchmark point, this work's value, and the public-code value. Benchmark~BM is $\mzero=45$~TeV, $\mhalf=1.3$~TeV, $\tb=5$, $A_0=0$. The last three rows are the dimension-five proton lifetime at the listed benchmark points, with $M_2=\mu=1$~TeV and $\phi_{13}=0$.  The two entries in the third column are the values computed by our {\tt Python} and {\tt Mathematica} codes, in that order.}
\label{tab:codebench}
\medskip
\begin{tabular}{@{}llll@{}}
\toprule
Observable & Benchmark & This work & Public code \\
\midrule
$m_h$ (fixed order)   & BM, CMSSM input      & ${\sim}142$~GeV                 & \texttt{SPheno}: $142.3$~GeV         \\
$m_h$ (resummed)      & BM                   & ${\approx}125$~GeV             & \texttt{FeynHiggs}: $124.7$~GeV      \\
$\Ohsq$ (higgsino)    & $\mu=1.1$~TeV        & $0.12$                         & \texttt{micrOMEGAs}: $0.122$         \\
$\sigSI$ (higgsino)   & $\mu\le1.1$~TeV      & $(2-3){\times}10^{-47}$~cm$^2$ & \texttt{micrOMEGAs}: $(2.2 - 2.6)\times10^{-47}$~cm$^2$    \\
$|\mu|$ (constrained-$\mu$) & $\tb=5$        & multi-TeV                      & \texttt{SPheno}: $7$--$33$~TeV       \\
$\tau_p(p\to K^+\bar\nu)$ [$10^{34}$~yr] & $(2,12)$ & $2.14879727$/$2.14879728$ & none available (see the text) \\
                     & $(5,20)$ & $3.78359687$/$3.78359684$ & \\
                     & $(8,25)$ & $2.88037069$/$2.88037071$ & \\
\bottomrule
\end{tabular}
\end{table*}

The free-$\mu$ scan thus leaves the survival wedge intact and adds one feature the universal model lacks, i.e., a light electroweakino, whose DM fate (a well-tempered strip and a higgsino branch) is taken up in \cref{sec:dm_freemu}.

\section{Lightest supersymmetric particle}\label{sec:dm}

Assuming exact $R$-parity symmetry, the LSP is stable, and its \emph{fate} is a test of the model: it must neither over-close the Universe nor be seen by a direct detection experiment. In our spectrum, the LSP is the lightest neutralino $\tilde\chi^0_1$, a weakly interacting massive particle (WIMP), whose gauge content (and hence its relic abundance and scattering rate) is fixed by the electroweakino sector $(M_1,M_2,\mu)$. That sector is realized, as discussed in \cref{sec:scan}, in two ways that give qualitatively different fates for the LSP: the constrained-$\mu$ scheme (\cref{sec:dm_cmu}) and the free-$\mu$ scheme (\cref{sec:dm_freemu}), where a light electroweakino (and hence a thermal WIMP candidate) reappears. We first set up the relic abundance and the direct detection cross section for comparison with the data, then take the two schemes in turn. Throughout, the LSP analysis lives in the electroweakino sector and leaves the $(\tb,\mzero)$ bounds of \cref{sec:walls} untouched.

\subsection{Relic density and direct detection}\label{sec:dm_tools}

We impose two DM conditions with the following data. First, if $\tilde\chi^0_1$ is a thermal relic, its abundance must not exceed the measured cold DM density $\Ohsq=0.120\pm0.001$~\cite{Planck:2018vyg}; we impose $\Ohsq\le0.122$, the one-sided $95\%$ confidence-level upper limit on the measured value, saturated when the neutralino is all of the DM. Because the squarks and sleptons here sit at tens of TeV and decouple, the abundance is fixed entirely by the electroweakino sector $(M_1,M_2,\mu)$ and their mixing. Thermal freeze-out sets it through the thermally averaged annihilation cross section $\langle\sigma v\rangle$,
\begin{equation}
\label{eq:relic}
\Ohsq \simeq 0.12\times\frac{2\times10^{-26}\,{\rm cm^3\,s^{-1}}}{\langle\sigma v\rangle}
~.
\end{equation}
A weakly coupled (small-$\langle\sigma v\rangle$) state over-closes the Universe. A pure bino, generic when $M_1$ is the smallest mass, has no $SU(2)_L$ or higgsino gauge coupling to annihilate through. So with only the very heavy sfermions for $t$-channel exchange, it over-closes the Universe, $\Ohsq\gg0.12$. In this case, efficient annihilation requires a higgsino (or wino) admixture, which depends on the scheme.

Second, a neutralino that makes up the DM also scatters off nuclei~\cite{Goodman:1984dc,Drukier:1986tm}; the strongest current bound is the 2024 LUX-ZEPLIN (LZ) spin-independent limit~\cite{LZ:2024zvo}, $\sigSI\lesssim2$--$3\times10^{-47}\,$cm$^2$ near $1$~TeV. The cross section is controlled by Higgs exchange and scales with the \emph{higgsino fraction} of $\tilde\chi^0_1$: a pure bino barely scatters, a pure higgsino more, and a \emph{mixed} bino--higgsino scatters \emph{strongly},
\begin{equation}
\label{eq:sigSI}
\sigSI
=
\frac{4}{\pi}\,\mu_{\rm red}^2\,f_p^2
~, \qquad
f_p
\propto
\frac{g_{\tilde\chi\tilde\chi h}}{m_h^2}
~, \qquad
g_{\tilde\chi\tilde\chi h}
\propto
N_{11}\,N_{1H}
~,
\end{equation}
where $\mu_{\rm red}$ is the neutralino--nucleon reduced mass and $g_{\tilde\chi\tilde\chi h}\propto N_{11}N_{1H}$ (bino $\times$ higgsino component) is largest for a comparably mixed state. Only the light-Higgs exchange survives: with the heavy-Higgs sector at the squark scale, we fix $M_A=20$~TeV, so the $H$ piece is suppressed by $\sim m_h^2/M_A^2\approx4\times10^{-5}$ and is negligible; the relic density is likewise far from any $s$-channel $A$-pole. The LSP makes up a fraction $f\equiv\min\!\big(\Ohsq/0.12,\,1\big)$ of the local DM density, up to $f = 1$, so the limit reads $f\,\sigSI\le\sigSI^{\rm LZ}$. An under-abundant relic ($f<1$) has its rate lowered below the raw $\sigSI$, whereas an over-abundant relic has $f=1$ and its raw $\sigSI$ is tested directly. The largest fraction it can constitute at a given mass is then $f_{\max}=\min\!\big(1,\,\sigSI^{\rm LZ}/\sigSI\big)$. The relic abundance and $\sigSI$ are computed with a Sommerfeld-corrected analytic freeze-out and a one-loop $\sigSI$, cross-checked against \texttt{micrOMEGAs}; the constrained-$\mu$ bino relic below is taken directly from \texttt{micrOMEGAs}.

\subsection{The constrained-\texorpdfstring{$\mu$}{mu} scheme}\label{sec:dm_cmu}

\begin{figure*}[htbp]
\centering
\begin{minipage}[t]{0.5\textwidth}\centering
\includegraphics[width=\linewidth]{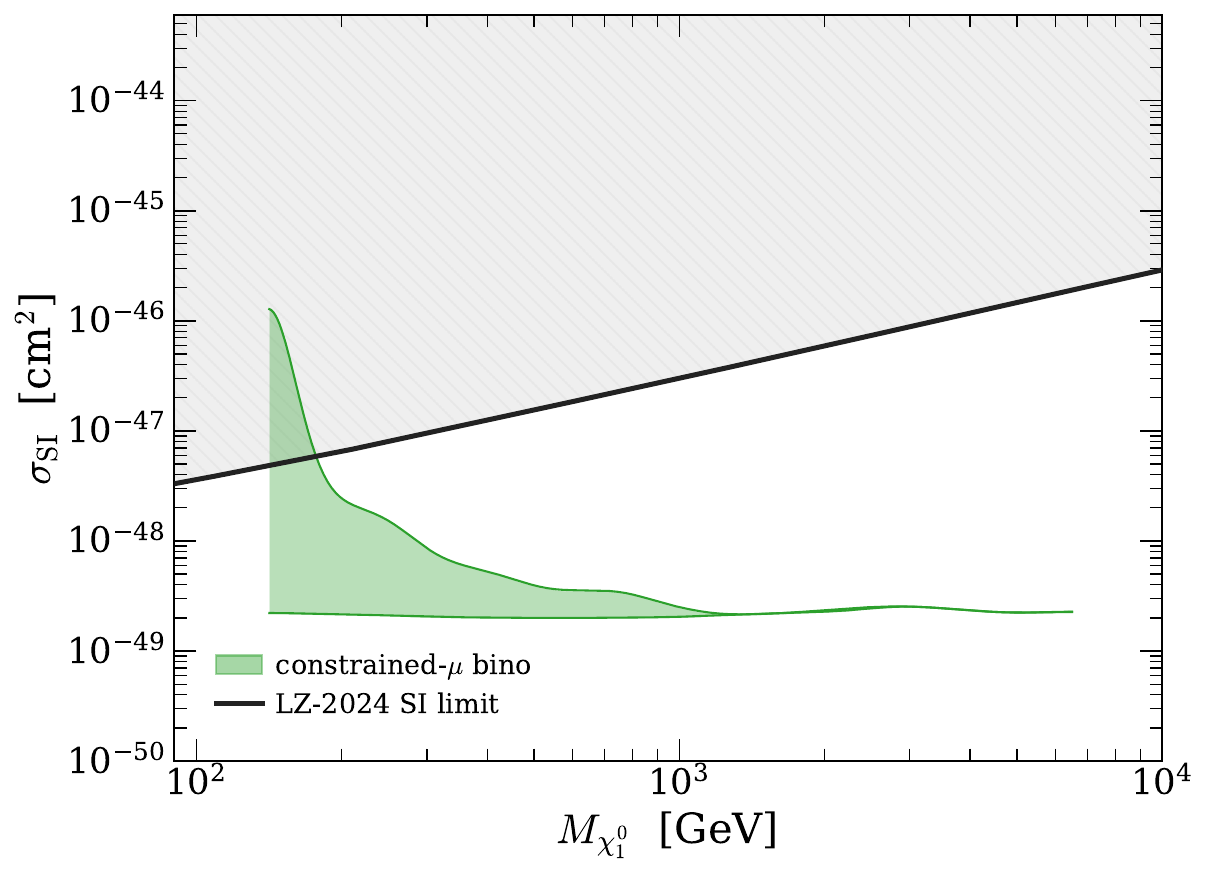}\\[2pt]
(a) constrained-$\mu$ scheme (universal model)
\end{minipage}\hfill
\begin{minipage}[t]{0.5\textwidth}\centering
\includegraphics[width=\linewidth]{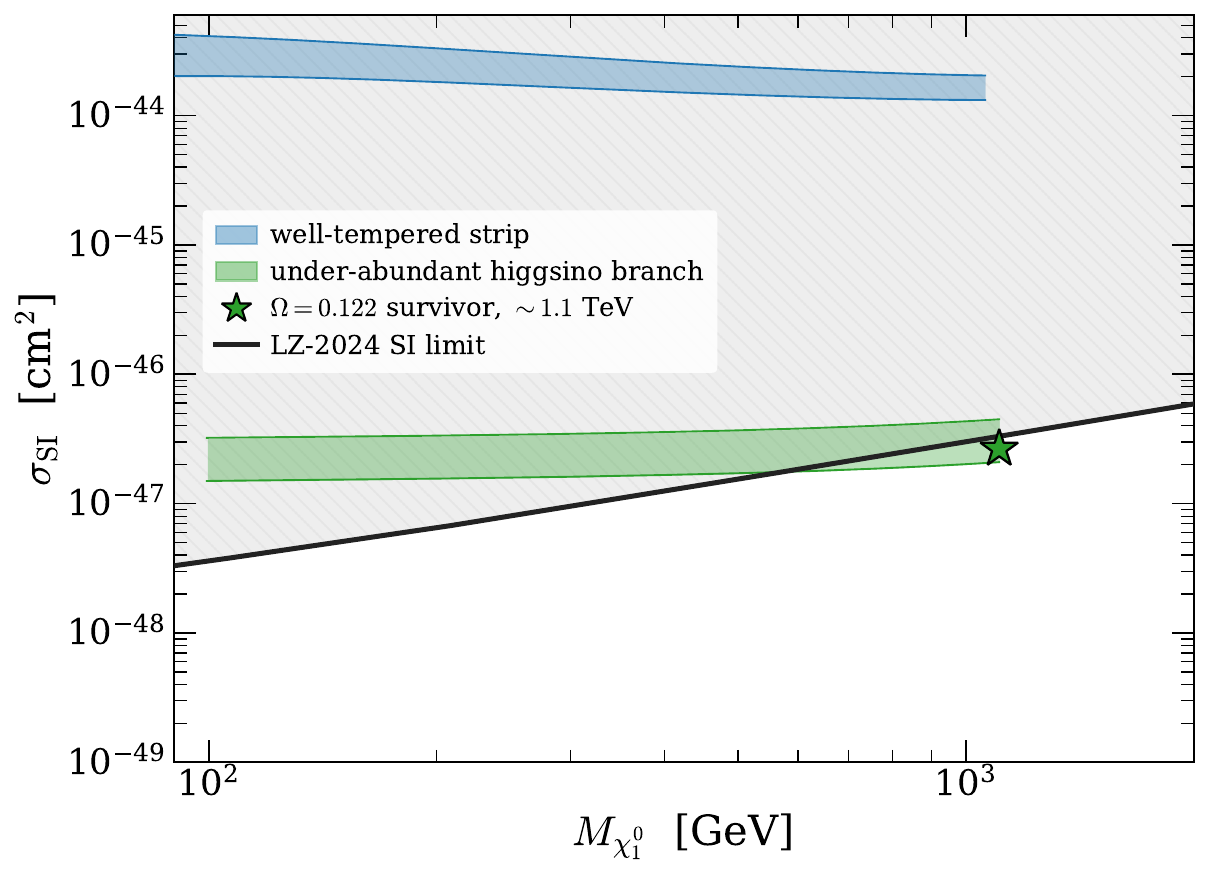}\\[2pt]
(b) free-$\mu$ scheme (NUHM extension)
\end{minipage}
\caption[The fate of the LSP: constrained-$\mu$ versus free-$\mu$.]{Fate of the LSP: spin-independent cross section with the nucleon as a function of $M_{\chi^0_1}$ against the LZ-2024 limit~\cite{LZ:2024zvo}, swept over the $\tb\in[2,8.0]$ and taking $M_A=20$~TeV. (a) Constrained-$\mu$ (the universal model): the bino (green), with $\mu=7$--$33$~TeV fixed by EWSB, is thermally \emph{over-abundant}, so its local density fraction is capped at $f=1$ and its \emph{raw} $\sigSI$ is what LZ tests. (b) Free-$\mu$ (the NUHM extension): the two strips are the \emph{same} neutralino with different compositions (dialed by $\mu/M_1$), running from the LEP chargino mass lower bound ($M_{\chi^0_1}\gtrsim100$~GeV) to the $\sim1.1$~TeV thermal-higgsino point where they merge (marked by the green star). The blue region is the well-tempered $\Ohsq=0.12$ strip ($f=1$, $\sigSI \simeq 10^2$--$10^3$ above LZ, excluded, not rescalable). The green region is the $\Ohsq\le0.122$ higgsino branch (raw $\sigSI$ touching/above LZ, but \emph{allowed} if under-abundant). Relic and $\sigSI$ are computed by \texttt{micrOMEGAs}.}
\label{fig:dm_dd}
\end{figure*}

When $\mu$ is fixed by radiative EWSB (\cref{sec:scan}), it is driven to the multi-TeV scalar scale, $|\mu|=7$--$33$~TeV. Therefore, $M_1<M_2\ll|\mu|$, and the LSP is a nearly pure bino ($\gtrsim99.9\%$ bino). Evaluated with \texttt{micrOMEGAs} along the constrained-$\mu$ surviving region, swept over the $\tb\in[2,8.0]$ in \cref{fig:dm_dd}(a), bino mass $M_{\chi^0_1}\simeq0.1$--$7$~TeV, sfermions at tens of TeV, the thermal relic is hugely \emph{over-abundant}:
\begin{equation}
\label{eq:cmu_relic}
\Ohsq \sim (10^{4}\text{--}10^{5})\times\,0.12 ~,
\end{equation}
because the tiny $\sim0.1\%$ higgsino admixture leaves only a feeble $\tilde\chi\tilde\chi\to W^+W^-,ZZ,hh$ annihilation through the $7$--$33$~TeV higgsino. As a standard thermal relic, the constrained-$\mu$ bino thus over-closes the Universe by some five orders of magnitude. The heavy $\mu$, the split $M_1<M_2$, and the tens-of-TeV sfermions leave no efficient annihilation channel. In this case, $\Ohsq\le0.122$ could be reached only through, for example, non-thermal dilution, after which it becomes a direct detection-safe subdominant relic.

Direct detection is mostly blind to it. Over most of the region, the LSP-nucleon cross section $\sigSI\simeq2$--$6\times10^{-49}\,$cm$^2$, two to three orders \emph{below} LZ (\cref{fig:dm_dd}(a)). The only exception is the high-$\tb$ ($\simeq8.0$), low-mass ($M_{\chi^0_1}\lesssim0.2$~TeV) corner, where the $\tb$-enhanced coupling lifts $\sigSI$ up to $\sim10^{-46}\,$cm$^2$ (reaching $8\times10^{-46}$ at the extreme). There, the bino cannot be $100\%$ of the DM and would have to be diluted to a subdominant fraction. Under exact $R$-parity, the relic excess is removed by a standard late entropy injection, a modulus or saxion decaying after bino freeze-out but before the big bang nucleosynthesis dilutes $\Ohsq$ by the required $\sim10^{5}$. After the dilution, the bino is a harmless, direct detection-safe subdominant relic and the observed $\Ohsq=0.12$ should be supplied by another sector (such as an axion). The same excess is equally avoided if the bino is not the true LSP: a lighter axino or gravitino, natural in the Peccei--Quinn or supergravity sectors that would also supply the axion, makes the bino a next-to-lightest state that decays into it. Consequently, the surviving abundance is set by the lighter LSP rather than the over-abundant bino. The constrained-$\mu$ scheme therefore offers no thermal WIMP DM candidate.

\subsection{The free-\texorpdfstring{$\mu$}{mu} scheme}\label{sec:dm_freemu}

If instead we relax the universal hypothesis in the Higgs sector, we have the NUHM extension of \cref{sec:scan_headline}, where $\mu$ becomes a free low-scale input while the sfermions stay universal.  In such a scheme, a light electroweakino and a thermal WIMP candidate reappear. The relic condition $\Ohsq=0.12$ is then met on two thin regions, drawn in \cref{fig:dm_dd}(b) as \emph{strips} because each is swept over the $\tb\in[2,8.0]$. Since $\sigSI$ carries a $\tb$-enhanced down-type Higgs coupling, a fixed neutralino mass broadens into a band. The two strips occupy the \emph{same} mass range but differ in the neutralino \emph{composition}, which is a third axis that the $(M_{\chi^0_1},\sigSI)$ projection hides and that is dialed by the ratio $\mu/M_1$.  For $\mu\simeq M_1$, the LSP is a well-tempered bino--higgsino mixture~\cite{Arkani-Hamed:2006wnf} (blue), while for $\mu\ll M_1$ (a heavy bino) it is a nearly pure higgsino (green). Both strips run from the LEP chargino bound on the left ($m_{\chi^\pm_1}>103.5$~GeV forces $M_{\chi^0_1}\gtrsim100$~GeV, since either configuration has a light chargino near the LSP mass) to the $\sim1.1$~TeV thermal-higgsino point on the right, where they merge.  Above $M_1\simeq1.1$~TeV no amount of mixing brings the bino down to $\Ohsq=0.12$, so it becomes the pure higgsino, which itself saturates $\Ohsq=0.12$ at $\mu\simeq1.1$~TeV. Heavier neutralinos over-close the Universe.

The two strips meet direct detection very differently. The well-tempered strip sits at $\Ohsq=0.12$ ($f=1$), but its $\mathcal{O}(1)$ higgsino admixture gives a raw $\sigSI\sim1.4\times10^{-44}\,$cm$^2$, $10^2$--$10^3$ above LZ. It is thus excluded by direct detection and, being already $100\%$ of the DM, cannot be rescued by density rescaling. The higgsino branch is instead under-abundant. For $|\mu|<M_1$, the pure higgsino has mass $\simeq|\mu|$ and $\Ohsq\simeq0.12\,(\mu/1.1\,\mathrm{TeV})^2<0.12$, the leading-order thermal scaling ($\Ohsq\propto1/\langle\sigma v\rangle$, $\langle\sigma v\rangle\propto1/\mu^2$) normalized to the $1.1$~TeV thermal higgsino (up to a mild logarithmic and Sommerfeld/co-annihilation correction folded into the calibration). Its raw $\sigSI\simeq2$--$3\times10^{-47}\,$cm$^2$ is close to the LZ bound and lies above it at low mass. Since the higgsino is only a fraction $f=\Ohsq/0.12=(\mu/1.1\,\mathrm{TeV})^2<1$ of the local density, the direct detection rate is $f\,\sigSI$, which drops below LZ across the whole branch. Hence, the higgsino can be a viable (generally subdominant) DM component: up to $100\%$ at the $\sim1.1$~TeV saturation (marked by the green star sitting right below the LZ bound) and a fraction $f$ below it. Every $\Ohsq\le0.122$ point is higgsino-dominated, and no bino-dominated LSP is ever under-abundant. This thermal-WIMP candidate is a feature of the NUHM extension.

\section{Discussions and outlook}\label{sec:wedge}

The surviving region of \cref{eq:wedge} is small and sharply bounded, and each of its edges is the target of a near-future experiment, as detailed below.

\paragraph{Proton decay.} The lower bound on
$\mzero$ is not a comfortable distance below the current limit: the lower edge sits at
$\tau_p(p\to K^+\bar\nu)\approx6.61\times10^{33}\,$yr, just above the present Super-Kamiokande
bound it was forced to clear~\cite{PDG26}. Hyper-Kamiokande~\cite{Hyper-Kamiokande:2018ofw}, with roughly an order-of-magnitude
greater sensitivity in this channel, will probe lifetimes into the $\sim\!10^{34}$--$10^{35}\,$yr
range and so test the low-$\mzero$ edge directly. The prediction is sharp: either
Hyper-Kamiokande (and DUNE in the complementary $K^+$ channel) sees the $K^+\bar\nu$ mode at a
rate consistent with $\mzero$ near the lower edge, or a null result pushes that edge upward, tightening the region toward smaller $\tb$, subject to the factor-few hadronic
uncertainty of \cref{sec:walls}.

\paragraph{Electroweakinos.} The mini-split pattern places the gauginos and higgsinos at $\sim$~TeV, far below the scalars, so they are the only directly accessible superpartners. The heavy Higgs states $H,A,H^\pm$ also sit at $\gtrsim m_0$, leaving the SM-like Higgs boson as the sole accessible scalar at low energies. In the constrained-$\mu$ scheme, the ordering of gauginos is fixed: a bino $\tilde\chi^0_1\simeq M_1$, a wino-like $\tilde\chi^\pm_1,\tilde\chi^0_2\simeq M_2\simeq 2M_1$ (the lighter chargino mass ranging up to a few TeV in the allowed region), and a decoupled higgsino ($|\mu|=7$--$33$~TeV). Since $M_1<M_2$, the winos decay promptly to the bino, giving the golden channel $pp\to\tilde\chi^\pm_1\tilde\chi^0_2\to W^{(*)}\tilde\chi^0_1\,(Z/h)\,\tilde\chi^0_1$, seen as a trilepton or $\ell\ell b\bar b$ final state with missing transverse energy. The high-luminosity LHC (HL-LHC) reaches wino-bino masses near $1$~TeV~\cite{CidVidal:2018eel}, and a $100$-TeV collider (FCC-hh)~\cite{FCC:2018vvp,Golling:2016gvc} extends the search to several TeV, covering the full electroweakino window predicted here.  In the free-$\mu$ (NUHM) scheme, the surviving $\sim1.1$~TeV higgsino is nearly degenerate and appears as soft leptons with a mono-jet or a disappearing track, out of HL-LHC pure-electroweak reach but a clean FCC-hh target.

\paragraph{Gluino.} A sharp signature is the gluino placed at $m_{\tilde g}\simeq(4$--$7)\,M_1$ by gaugino-mass unification, a few TeV when the electroweakinos are light. Because the squarks are very heavy, $\tilde g\to q\bar q\,\tilde\chi$ proceeds only through an off-shell squark of mass $\sim m_0$, with the width scaling as $m_{\tilde g}^5/m_0^4$ and leaving the gluino long-lived, a hallmark of split spectra. Taking $m_0\sim45$~TeV and $m_{\tilde g}\sim3$~TeV as an example, the decay length is macroscopic, and the leading probes are long-lived particle searches (R-hadrons tagged by anomalous ionization and time-of-flight, and stopped-gluino searches), with prompt multijet-plus-missing-energy when the decay is fast. The HL-LHC reaches $m_{\tilde g}\sim2.5$--$3$~TeV~\cite{CidVidal:2018eel} and FCC-hh $\sim10$--$15$~TeV~\cite{FCC:2018vvp,Golling:2016gvc}, spanning the entire allowed range predicted here. The model thus predicts a compact electroweakino system with a possibly long-lived multi-TeV gluino. Therefore, a $\sim$~TeV electroweakino or R-hadron signal with no accompanying light squarks or sleptons would point directly to the mini-split spectrum and select the light-$M_{1/2}$ corner.

\paragraph{Dark matter.} The free-$\mu$ DM target of \cref{sec:dm}, a $\sim1.1$~TeV nearly pure higgsino (a feature of the NUHM extension, not of the universal model, whose own LSP is an over-abundant bino requiring dilution), is a clean, fixed-mass prediction (relaxed to a higgsino branch $\mu\lesssim1.1$~TeV if one requires only $\Ohsq\le0.122$; \cref{sec:dm}). Its spin-independent cross section lies below the current LZ bound but
within the projected sensitivity of the next-generation multi-ton xenon experiments
(XLZD/DARWIN), which will cover essentially the entire higgsino-mass window, making the higgsino
DM a falsifiable hypothesis.

\paragraph{Electric dipole moments.} With strictly real, universal
soft terms, the electric dipole moments vanish (with only the SM CKM,
$d_e\sim10^{-38}\,e\,$cm), so they do not constrain the baseline scan~\cite{Roussy:2022cmp,Abel:2020pzs}. The
heavy-scalar/light-gaugino corner is nonetheless sensitive: a two-loop Barr--Zee contribution
survives the $1/\mzero^2$ decoupling of the one-loop sfermion piece.  Hence, even a small residual
$CP$-violating phase in the gaugino sector would generate an electron EDM within reach of
next-generation searches.  It is tight enough that the present electron-EDM limit~\cite{Roussy:2022cmp}
already requires roughly $\sin\varphi\lesssim10^{-3}$. The next round of electron- and neutron-EDM
experiments therefore probes the $CP$ structure of the light-gaugino corner: a nonzero EDM at
their sensitivity would point to this regime, while continued nulls tighten the bound on
gaugino-sector phases.

Taken together, these four probes reach the relevant sensitivity within the coming
decade: proton decay tests the lower edge, colliders the electroweakinos, direct detection the
higgsino, and EDMs the $CP$ structure.

\section{Conclusions}\label{sec:concl}

We have delineated the allowed parameter space of the minimal supersymmetric $\SO{10}$ with a
universal soft spectrum by requiring, on every point of a dense scan over
$\{\mzero, M_2, \mu, \tb, X_t, \MHC\}$, that all constraints hold at once. The constraints span four
physically distinct categories: theory (gauge coupling unification and vacuum stability), the robust
experimental limits (proton lifetime limit, $m_h = 125$~GeV, and the LEP chargino bound), dark matter data, and
the flavor and precision observables. The model is not excluded, but, under a single effective colored Higgs scale for the proton decay normalization and the \texttt{SPheno}-based constrained-$\mu$ treatment, it survives only in a sharply bounded, low-to-moderate-$\tb$ mini-split region: multi-tens-of-TeV scalars,
$\mzero \gtrsim 11.0$~TeV for a dressing electroweakino mass $m_\chi=1$~TeV and $\tb \lesssim 8.0^{+5.5}_{-2.0}$, subject to the systematic uncertainty in the EFT-resummed calculation of $m_h$, above $\sim$~TeV gauginos, with the colored Higgs mass
pinned by unification to $\MHC \in [0.58, 1.43] \times 10^{16}$~GeV.

Proton decay inverts the usual decoupling intuition into a \emph{lower} bound on $\mzero$ that rises with $\tb$. The Higgs mass sets an upper bound that descends with $\tb$.  These two constraints meet near $\tb \approx 8.0$ (best estimate) with no extra assumption. Unification ties the proton rate to the colored Higgs mass
through $\tau_p \propto \MHC^2$, closing, within the single effective-$\MHC$ parametrization adopted here, the obvious escape of raising $\MHC$: a lighter scalar cannot be traded for a heavier triplet without spoiling unification. In a complete color triplet sector, where the dimension-five decay depends on the inverse triplet mass matrix and unification on threshold combinations, this is a model assumption. We estimate the resulting decoupling of the unification-effective and decay-effective triplet scales to be a factor $\lesssim2$ on $\MHC$, hence $\lesssim1.5$ on the $\mzero$ bound, for a color triplet mass spread up to an order of magnitude (\cref{eq:mhc_decouple}), barring a tuned cancellation in the inverse mass matrix.

Solving the electroweak symmetry-breaking conditions point by point, rather than treating $M_2$
and $|\mu|$ as free parameters, sharpens this picture. The three edges above depend at leading order only on
$\{\mzero, \MHC, \tb\}$, so the lower bound and the $\tb$ upper bound are unchanged. What the constrained-$\mu$
treatment removes is the light electroweakino, since radiative symmetry breaking drives
$|\mu|$ to the multi-TeV scalar scale, well above $M_2$ at every surviving point.  Therefore, the lightest
electroweakino is the bino (since $M_1<M_2$) and no light higgsino is realized.

Dark matter leaves the $(\tb,\mzero)$ bounds untouched but sharply constrains the fate of the lightest supersymmetric particle. In the universal model, the lightest neutralino is a bino and radiative electroweak symmetry breaking drives $|\mu|$ heavy, so its thermal relic is hugely over-abundant yet invisible to direct detection: under exact $R$-parity, a late entropy dilution is required to render $\Ohsq\le0.122$, with the observed density supplied by another sector (\cref{sec:dm_cmu}). A light thermal-relic neutralino reappears only in the minimal non-universal-Higgs-mass extension in which $\mu$ is a free parameter; there LUX-ZEPLIN direct detection~\cite{LZ:2024zvo} excludes the entire well-tempered bino--higgsino strip and collapses the viable candidate to a single $\sim\!1.1$~TeV nearly pure higgsino. Because this acts orthogonally to the $(\tb,\mzero)$ plane, the allowed region and its bounds are unaffected. Equivalently, relaxing the requirement that the lightest neutralino be the dark matter or that it be a thermal relic at all does not qualitatively alter the mini-split conclusion, since the bounds $\mzero\gtrsim11.0$~TeV (again at $m_\chi=1$~TeV) and $\tb\lesssim8.0$ rest entirely on the proton decay limit, the Higgs mass, and gauge unification. The flavor and precision observables ($\mu\to e\gamma$, $(g-2)_\mu$, and electric dipole moments) are also checked to be non-binding: the $Y_D$-induced slepton mixing behind $\mu \to e\gamma$ falls below MEG~II once $\mzero \gtrsim 5$~TeV, comparable to the heaviness the proton lifetime lower bound already demands.

In short, the minimal supersymmetric $\SO{10}$, if realized at all, is pushed to the cornered mini-split and, as detailed in \cref{sec:wedge}, testable in the coming decades, most directly through Hyper-Kamiokande proton decay and electroweakino searches at the HL-LHC and a future $100$-TeV collider, with dark matter direct detection and electric dipole moment experiments covering parts of the region.

\acknowledgments

This work was supported in part by the National Science and Technology Council under Grant 
No.~NSTC114-2112-M-002-020-MY3 (CWC) and in part by Grant-in-Aid for Scientific Research from the Ministry of Education, Culture, Sports, Science and Technology, Japan (No.~25H00653).
Numerical calculations in this work were done in collaboration with {\tt Claude-Code Opus 4.8} under our supervision, including writing the code, running other HEP packages, making plots and tables, and part of the writing.

\providecommand{\href}[2]{#2}\begingroup\raggedright\endgroup

\end{document}